\documentclass[12pt]{article}
\usepackage{epsfig}
\usepackage{cite}

\newcommand{\mysection}{\setcounter{equation}{0}\section}

\def\beq{\begin{equation}}
\def\eeq{\end{equation}}
\def\beqa{\begin{eqnarray}}
\def\eeqa{\end{eqnarray}}
 
\newlength{\dinwidth} \newlength{\dinmargin}
\begin{document}

\begin{center}
{\Large \bf Soft anomalous dimensions and resummation in QCD}
\end{center}
\vspace{2mm}
\begin{center}
{\large Nikolaos Kidonakis}\\
\vspace{2mm}
{\it Department of Physics, Kennesaw State University,\\
Kennesaw, GA 30144, USA}
\end{center}
 
\begin{abstract}
I discuss and review soft anomalous dimensions in QCD that describe soft-gluon threshold resummation for a wide range of hard-scattering processes. The factorization properties of the cross section in moment space and renormalization-group evolution are implemented to derive a general form for differential resummed cross sections. Detailed expressions are given for the soft anomalous dimensions at one, two, and three loops, including some new results, for a large number of partonic processes involving top quarks, electroweak bosons, Higgs bosons, and other particles in the Standard Model and beyond.

\end{abstract}
 
\mysection{Introduction}

This review discusses soft anomalous dimensions that control soft-gluon threshold resummation in QCD. Resummation follows from factorization properties of the cross section \cite{GS87,CT89,NKGS1,CLS97,NKGS2,KOS1,KOS2,LOS,ADS,HRSV,MFNK} and it provides a formalism for calculating contributions to higher-order corrections that are theoretically important and numerically significant. These soft-gluon contributions take the form of logarithms of a variable that is a measure of the available energy for additional radiation in a process. 

Beyond leading logarithms, resummation is crucially dependent on the color exchange in the hard scattering. One-loop calculations of soft anomalous dimensions are necessary to achieve next-to-leading-logarithm (NLL) accuracy while two-loop calculations are needed for next-to-next-to-leading-logarithm (NNLL) accuracy. The current state of the art is three-loop calculations which are needed for  next-to-next-to-next-to-leading-logarithm (N$^3$LL) accuracy.

Soft-gluon resummation is particularly relevant near partonic threshold and it has been used for a large number of hard-scattering processes. In fact, soft-gluon resummation is often relevant even far from threshold for many Standard Model (SM) and Beyond the Standard Model (BSM) processes. There is a vast number of results in the literature that have been provided using various approaches, schemes, gauges, and definitions. Therefore, in addition to reviewing past results, it is useful to provide a comprehensive and unified treatment using a common formalism and notation for all these results. In this review all results are shown using the standard moment-space resummation formalism in QCD for, in general differential, cross sections in the $\rm {\overline {MS}}$ scheme and in Feynman gauge. The expressions for the resummed cross section are shown in single-particle-inclusive (1PI) kinematics, but the soft anomalous dimensions are the same in other kinematics choices, such as pair-invariant-mass (PIM) kinematics.

This paper is meant as a focused review of theoretical work on resummation and as a compendium of results on soft anomalous dimensions and other quantities in the resummed expressions. It is not a review of phenomenological papers with applications of resummations; there are many hundreds of such papers, and a recent review for many processes involving top quarks can be found in Ref. \cite{NKtoprev}. Furthermore, this is a review on standard moment-space QCD resummation; it is not a review of related work using alternative approaches such as soft-collinear effective theory (SCET) or non-relativistic QCD (NRQCD). A comparative review of resummation approaches in QCD and SCET was given in Ref. \cite{NKBP}. A main goal of this paper is to provide a common terminology and notation for the large number of results in the literature for a large variety of SM and BSM processes, in the hope that it will be useful in comparing past results and in future applications.

In Section 2 we provide the formalism for soft-gluon resummation based on factorization and renormalization-group evolution. Section 3 presents fixed-order expansions of resummed cross sections at next-to-leading order (NLO), next-to-next-to-leading order (NNLO), and next-to-next-to-next-to-leading order (N$^3$LO). Section 4 has results at one, two, and three loops for the cusp anomalous dimension for the separate cases of two massive lines, or two massless lines, or one massive and one massless lines. The soft anomalous dimensions for many processes with trivial color structure are discussed in Section 5, for high-$p_T$ electroweak-boson production and related processes in Section 6, for single-top production and related processes in Section 7, for top-antitop production and related processes in Section 8, for jet production and related processes in Section 9, and for several $2 \to 3$ processes in Section 10. A concluding summary is given in Section 11.

\mysection{Soft-gluon resummation}

In this section we briefly review the moment-space QCD soft-gluon resummation formalism. We discuss the factorization and refactorization of the cross section, renormalization-group evolution (RGE), the eikonal approximation, soft anomalous dimensions, and the resummed cross section. For simplicity we discuss $2 \to 2$ processes but also explain how this generalizes to $2 \to n$ processes \cite{MFNK}. For specificity we choose 1PI kinematics but also discuss modifications for PIM kinematics.

\subsection{Factorization, RGE, and resummation}

The factorized form of the, in general, differential cross section $d\sigma_{AB \rightarrow 12}$ in hadronic collisions for the process $AB \rightarrow 12$ is 
\beq
d\sigma_{AB \rightarrow 12}=\sum_{a,b} \; 
\int dx_a \, dx_b \,  \phi_{a/A}(x_a, \mu_F) \, \phi_{b/B}(x_b, \mu_F) \, 
d{\hat \sigma}_{ab \rightarrow 12}(\mu_F, \mu_R) \, ,
\label{facphi}
\eeq
where $\mu_F$ is the factorization scale, $\mu_R$ is the renormalization scale, $\phi_{a/A}$ ($\phi_{b/B}$) are parton distribution functions (pdf) for parton $a$ ($b$) in hadron $A$ ($B$), and $d{\hat \sigma}_{ab \rightarrow 12}$ is the differential hard-scattering partonic cross section.

We consider partonic processes $ab\rightarrow 12$ with 4-momenta $p_a +p_b \rightarrow p_1 +p_2$, and define the usual kinematical variables $s=(p_a+p_b)^2$, $t=(p_a-p_1)^2$, and $u=(p_b-p_1)^2$. In 1PI kinematics we choose, without loss of generality, particle 1 as the observed particle. We also define the threshold variable $s_4=s+t+u-m_1^2-m_2^2$ where the masses $m_1$ and $m_2$ can be zero or finite. As we approach partonic threshold, with vanishing energy for additional radiation, we have $s_4 \rightarrow 0$. If an additional gluon with momentum $p_g$ is emitted in the final state, then we can equivalently write $s_4=(p_2+p_g)^2-m_2^2$, so as $p_g$ goes to 0 (i.e. we have a soft gluon), we again see that $s_4$ describes the extra energy in the soft emission and that $s_4 \rightarrow 0$. We note that we can extend our formulas to the general case of multi-particle final states, i.e. $2 \to n$ processes \cite{MFNK}, by replacing $m_2^2$ by $(p_2+\cdots+p_n)^2$ in the expressions. 

With the incoming partons $a$ and $b$ arising from hadrons (e.g. protons/antiprotons) $A$ and $B$, we define the hadron-level variables $S=(p_A+p_B)^2$, $T=(p_A-p_1)^2$, $U=(p_B-p_1)^2$, and $S_4=S+T+U-m_1^2-m_2^2$. Writing $p_a=x_a p_A$ and $p_b=x_b p_B$, where $x_a$ and $x_b$ are the fractions of the momenta carried by, respectively, partons $a$ and $b$ in hadrons $A$ and $B$, we have the relations $s=x_a x_b S$, $t=x_aT+(1-x_a)m_1^2$, and $u=x_bU+(1-x_b)m_1^2$. Using the above relations, we find that 
\beq
\frac{S_4}{S}=\frac{s_4}{s}-(1-x_a)\frac{\left(u-m_2^2\right)}{s}-(1-x_b)\frac{\left(t-m_2^2\right)}{s}+(1-x_a)(1-x_b) \frac{\left(m_1^2-m_2^2\right)}{s} \, .
\label{S4}
\eeq
The last term in the above equation can be ignored near threshold, in the limit $x_a\rightarrow 1$ and $x_b \rightarrow 1$, since it involves the product $(1-x_a)(1-x_b)$.

We next discuss the factorization of the cross section in integral transform space \cite{NKGS1,NKGS2,LOS}. We define Laplace transforms (shown with a tilde) of the partonic cross section as 
$d{\tilde{\hat\sigma}}_{ab \rightarrow 12}(N)=\int_0^s (ds_4/s) \,  e^{-N s_4/s} \, d{\hat\sigma}_{ab \rightarrow 12}(s_4)$, with transform variable $N$, and we also define the transforms of the pdf as ${\tilde \phi}(N)=\int_0^1 e^{-N(1-x)} \phi(x) \, dx$. We note that under transforms the logarithms of $s_4$ in the perturbative series produce logarithms of $N$, and we will show that the latter exponentiate.

We then consider the parton-parton cross section $d\sigma_{ab \rightarrow 12}$, which has the same form as Eq. (\ref{facphi}) but with incoming partons instead of hadrons \cite{NKGS1,CLS97,NKGS2,KOS1,KOS2,LOS}  
\beq
d\sigma_{ab \rightarrow 12}(S_4)=
\int dx_a \, dx_b \,  \phi_{a/a}(x_a) \, \phi_{b/b}(x_b) \, 
d{\hat \sigma}_{ab \rightarrow 12}(s_4) \, ,
\label{factphi}
\eeq
noting that the leading power as $s_4 \rightarrow 0$ comes entirely from the flavor-diagonal distributions $\phi_{a/a}$ and $\phi_{b/b}$ \cite{NKGS2,KOS1}, and we define its transform as 
\beq
d{\tilde \sigma}_{ab \rightarrow 12}(N)=\int_0^S 
\frac{dS_4}{S} \,  e^{-N S_4/S} \, d\sigma_{a b \rightarrow 12}(S_4) \, . 
\label{cstr}
\eeq 

Using Eq. (\ref{S4}) (without the last term, which vanishes near threshold), we can rewrite the transform of the parton-parton cross section as  
\beqa
d{\tilde \sigma}_{ab \rightarrow 12}(N) &=& \int_0^1 dx_a e^{-N_a (1-x_a)} 
\phi_{a/a}(x_a) \int_0^1 dx_b e^{-N_b (1-x_b)} \phi_{b/b}(x_b)
\int_0^s \frac{ds_4}{s} e^{-N s_4/s} d{\hat \sigma}_{ab \rightarrow 12}(s_4)
\nonumber \\ 
&=& {\tilde \phi}_{a/a}(N_a) \, {\tilde \phi}_{b/b}(N_b) \, 
d{\tilde{\hat \sigma}}_{ab \rightarrow 12}(N) \, ,
\label{fac}
\eeqa
where $N_a=N(m_2^2-u)/s$ and $N_b=N(m_2^2-t)/s$ in 1PI kinematics (while $N_a=N_b=N$ in the corresponding formula in PIM kinematics).

Next, we introduce a refactorization of the cross section in terms of new functions $H_{ab\rightarrow 12}$, $S_{ab\rightarrow 12}$, $\psi_{a/a}$, $\psi_{b/b}$, $J_1$, and $J_2$ \cite{NKGS1,CLS97,NKGS2,KOS1,KOS2,LOS}. The process-dependent hard function $H_{ab\rightarrow 12}$ is purely short-distance, nonradiative, and infrared safe, and it comprises contributions from the amplitude and its complex conjugate. The soft function $S_{ab \rightarrow 12}$ describes the emission of noncollinear soft gluons and is also process-dependent. The coupling of soft gluons to the partons in the hard scattering processes is described by eikonal (Wilson) lines as ordered exponentials of the gauge field. The hard and soft functions are in general Hermitian matrices in color-exchange space, and the refactorized cross section involves the trace of their product. The functions $\psi_{a/a}$ and $\psi_{b/b}$ differ from the pdf $\phi_{a/a}$ and $\phi_{b/b}$, and they describe the dynamics of collinear emission from the incoming partons \cite{GS87,NKGS1,CLS97,NKGS2,KOS1,KOS2,LOS}. The functions $J_1$ and $J_2$ describe collinear emission from any final-state massless colored particles, and are absent otherwise. The refactorized form of the cross section \cite{NKGS1,NKGS2,KOS1,LOS} is then 
\beqa
d{\sigma}_{ab \rightarrow 12}&=&\int dw_a \, dw_b \, dw_1 \, dw_2 \, dw_S \, \psi_{a/a}(w_a) \, \psi_{b/b}(w_b) \, J_1(w_1) \, J_2(w_2) 
\nonumber \\ && \hspace{-31mm}\times
{\rm tr} \left[H_{ab\rightarrow 12}\left(\alpha_s(\mu_R)\right) \, 
S_{ab \rightarrow 12}\left(\frac{w_S \sqrt{s}}{\mu_F} \right)\right] \; 
\delta\left(\frac{S_4}{S}+w_a\frac{(u-m_2^2)}{s}
+w_b \frac{(t-m_2^2)}{s}-w_S -w_1-w_2\right)
\label{refact}
\eeqa
where the $w$'s are dimensionless weights, with $w_a$ and $w_b$ for $\psi_{a/a}$ and $\psi_{b/b}$, respectively, $w_1$ and $w_2$ for $J_1$ and $J_2$, respectively, and $w_S$ for $S_{ab\rightarrow 12}$. The argument in the delta function of Eq. (\ref{refact}) arises from the recasting of Eq. (\ref{S4}) in terms of the new weights, as
\beqa
\frac{S_4}{S}& = & -(1-x_a)\frac{\left(u-m_2^2\right)}{s}-(1-x_b)
\frac{\left(t-m_2^2\right)}{s}+\frac{s_4}{s}
\nonumber \\ 
&=& -w_a \frac{(u-m_2^2)}{s}- w_b \frac{(t-m_2^2)}{s}+w_S +w_1+w_2 \, .
\label{ws}
\eeqa
We note that $w_a \neq 1-x_a$ and $w_b \neq 1-x_b$ because they refer to different functions.

After taking a transform of Eq. (\ref{refact}), we have
\beqa
d{\tilde \sigma}_{ab \rightarrow 12}(N)\!\!\!\!\!&=&\!\!\!\!\! 
\int_0^1 dw_a e^{-N_a w_a} \psi_{a/a}(w_a) \int_0^1 dw_b e^{-N_b w_b} \psi_{b/b}(w_b)
\int_0^1 dw_1 e^{-N w_1} J_1(w_1)  \int_0^1 dw_2 e^{-N w_2} J_2(w_2)
\nonumber \\ && \!\!\! \times
{\rm tr}\left[H_{ab \rightarrow 12}\left(\alpha_s(\mu_R)\right) \int_0^1 dw_s e^{-N w_s}  
S_{ab \rightarrow 12}\left(\frac{w_s\sqrt{s}}{\mu_F} \right)\right] 
\nonumber \\ &=& 
{\tilde \psi}_{a/a}(N_a) \, {\tilde \psi}_{b/b}(N_b) {\tilde J_1}(N) {\tilde J_2}(N) \, \, 
{\rm tr} \left[H_{ab \rightarrow 12}\left(\alpha_s(\mu_R)\right) \, 
{\tilde S}_{ab \rightarrow 12}\left(\frac{\sqrt{s}}{N \mu_F} \right)\right] \, .
\label{refac}
\eeqa
All $N$-dependence has now been absorbed into the functions ${\tilde S}$, ${\tilde \psi}$, and ${\tilde J}$.

By comparing Eqs. (\ref{fac}) and (\ref{refac}), we find an expression for the hard-scattering partonic cross section in transform space,
\beq
d{\tilde{\hat \sigma}}_{ab \rightarrow 12}(N)=
\frac{{\tilde \psi}_a(N_a) \, {\tilde \psi}_b(N_b) \, {\tilde J_1} (N) \, {\tilde J_2} (N)}
{{\tilde \phi}_{a/a}(N_a) \, {\tilde \phi}_{b/b}(N_b)} \, \,  
{\rm tr} \left[H_{ab\rightarrow 12}\left(\alpha_s(\mu_R) \right) \, 
{\tilde S}_{ab \rightarrow 12}\left(\frac{\sqrt{s}}{N \mu_F} \right)\right] \, .
\label{sigN}
\eeq

We resum the $N$-dependence of the soft matrix ${\tilde S}^{ab \rightarrow 12}$ via renormalization-group evolution \cite{NKGS1,NKGS2}, 
\beq
{\tilde S}^b_{ab\rightarrow 12}=Z_{ab \rightarrow 12}^{\dagger} \; {\tilde S}_{ab \rightarrow 12} \; Z_{ab\rightarrow 12}
\label{Zab}
\eeq
where ${\tilde S}^b_{ab\rightarrow 12}$ is the bare quantity 
and $Z_{ab\rightarrow 12}$ is a (in general, matrix of) renormalization constant(s).
Thus, we have the renormalization group equation for ${\tilde S}_{ab \rightarrow 12}$,
\beq
\mu_R\frac{d{\tilde S_{ab\rightarrow 12}}}{d\mu_R}=\left(\mu_R \frac{\partial}{\partial \mu_R}
+\beta(g_s, \epsilon)\frac{\partial}{\partial g_s}\right) {\tilde S}_{ab\rightarrow 12}
=-\Gamma_{\!\! S \, ab\rightarrow 12}^{\dagger} \; {\tilde S}_{ab\rightarrow 12}
-{\tilde S}_{ab\rightarrow 12} \; \Gamma_{\!\! S \, ab\rightarrow 12}
\label{rges}
\eeq
where $\Gamma_{\!\! S \, ab\rightarrow 12}$ is the soft anomalous dimension (matrix), $g_s^2=4\pi\alpha_s$, and $\beta(g_s, \epsilon)=-g_s\epsilon/2+\beta(g_s)$ in $4-\epsilon$ dimensions where $\beta(g_s)=\mu_R\, dg_s/d\mu_R$ is the QCD beta function. We can also define the beta function in an alternative form in terms of $\alpha_s$ as 
\beq
\beta(\alpha_s)=\frac{d\ln\alpha_s}{d\ln\mu_R^2}=-\sum_{n=0}^{\infty}\beta_n \left(\frac{\alpha_s}{4\pi}\right)^{n+1} \, ,
\eeq
with $\beta_0=(11C_A-2n_f)/3$ \cite{GW,HDP}, where $C_A=N_c$ with $N_c$ the number of colors, and $n_f$ is the number of light quark flavors, $\beta_1=34 C_A^2/3-2 C_F n_f-10 C_A n_f/3$\cite{beta1a,beta1b,beta1c}, where $C_F=(N_c^2-1)/(2N_c)$, and \cite{beta2a,beta2b}
\beq
\beta_2=\frac{2857}{54}C_A^3+\left(C_F^2-\frac{205}{18}C_F C_A
-\frac{1415}{54}C_A^2\right)n_f
+\left(\frac{11}{9}C_F+\frac{79}{54}C_A\right)n_f^2 \, .
\eeq
Also $\beta_3$ \cite{beta3} is
\beqa
\beta_3&=&C_A^4 \left(\frac{150653}{486}-\frac{44}{9} \zeta_3 \right)
+C_A^3 n_f \left(-\frac{39143}{162}+\frac{68}{3} \zeta_3 \right)
+C_A^2 C_F n_f \left(\frac{7073}{486} - \frac{328}{9} \zeta_3 \right)
\nonumber \\ && \hspace{-3mm}
{}+C_A C_F^2 n_f \left(-\frac{2102}{27} + \frac{176}{9} \zeta_3 \right)
+23 C_F^3 n_f 
+C_A^2 n_f^2 \left(\frac{3965}{162}+\frac{56}{9} \zeta_3 \right)
+C_F^2 n_f^2 \left(\frac{338}{27}-\frac{176}{9} \zeta_3 \right)
\nonumber \\ && \hspace{-3mm}
{}+C_A C_F n_f^2 \left(\frac{4288}{243}+\frac{112}{9} \zeta_3 \right)
+\frac{53}{243} C_A  n_f^3 +\frac{154}{243} C_F n_f^3  
+\frac{d^{abcd}_A d^{abcd}_A}{N_A} \left(-\frac{80}{9}+\frac{704}{3} \zeta_3 \right)
\nonumber \\ && \hspace{-3mm}
{}+\frac{d^{abcd}_F d^{abcd}_A}{N_A} n_f \left(\frac{512}{9}
-\frac{1664}{3} \zeta_3 \right)
+\frac{d^{abcd}_F d^{abcd}_F}{N_A} n_f^2  \left(-\frac{704}{9}+\frac{512}{3} \zeta_3 \right)
\eeqa
where $\zeta_3=1.202056903\cdots$, $N_A=N_c^2-1$, $d_A^{abcd} d_A^{abcd}/N_A=N_c^2(N_c^2+36)/24$, $d_F^{abcd} d_A^{abcd}/N_A=N_c(N_c^2+6)/48$, and  
$d_F^{abcd} d_F^{abcd}/N_A=(N_c^4-6N_c^2+18)/(96 N_c^2)$.
The result for $\beta_4$ is given by a long expression in Ref. \cite{beta4}, and for QCD with $N_c=3$ it has the approximate numerical value $\beta_4 \approx 537148 - 186162 \, n_f + 17567.8 \, n_f^2 - 231.28 \, n_f^3 - 1.8425 \, n_f^4$.

\subsection{Eikonal approximation and soft anomalous dimensions}

From Eq. (\ref{rges}) we see that the evolution of the soft function is controlled by the soft anomalous dimension, $\Gamma_{\!\! S \, ab\rightarrow 12}$, which is also in general a matrix. In calculating $\Gamma_{\!\! S \, ab\rightarrow 12}$, we use the eikonal approximation, where the Feynman rules for diagrams with soft gluon emission 
simplify. For example in the emission of a soft gluon with four-momentum $k^{\mu}$ from a quark with final four-momentum $p^{\mu}$, we have the simplification 
\beq
{\bar u}(p) \, (-i g_s T^c) \, \gamma^{\mu}
\frac{i (p\!\!/+k\!\!/+m)}{(p+k)^2-m^2+i\epsilon} 
\rightarrow {\bar u}(p)\,  g_s T^c \, \gamma^{\mu}
\frac{p\!\!/+m}{2p\cdot k+i\epsilon}
={\bar u}(p)\, g_s T^c \,
\frac{v^{\mu}}{v\cdot k+i\epsilon}
\eeq
where ${\bar u}$ is a Dirac spinor, $T^c$ are the generators of SU(3), and 
$p^{\mu}=(\sqrt{s}/2) v^{\mu}$, with $v^{\mu}$ a four-velocity.
Thus, a typical one-loop diagram involving eikonal lines $i$ and $j$ is of the form 
\beq
g_s^2 \int \frac{d^n k}{(2 \pi)^n}  \frac{v_i^\mu}{(v_i \cdot k +i\epsilon)}  \frac{(-i) g_{\mu \nu}}{k^2} \frac{v_j^\nu}{(v_j \cdot k+i\epsilon)}\, ,
\eeq
which contributes a UV pole
$-(1/\epsilon)(\alpha_s/\pi) \, \theta \, \coth\theta$ where
$\theta=\ln[(v_i \cdot v_j+\sqrt{(v_i \cdot v_j)^2-v_i^2 v_j^2})/\sqrt{v_i^2 v_j^2}]$ is the cusp angle that we will discuss further in Sec. 4. For massless eikonal lines this expression for the UV pole simplifies to 
$-(1/\epsilon)(\alpha_s/\pi) \ln(2 v_i \cdot v_j/\sqrt{v_i^2 v_j^2})$.

The calculation of $\Gamma_{\!\! S \, ab\rightarrow 12}$ requires determining the coefficients of the ultraviolet poles of relevant eikonal diagrams \cite{NKGS1,NKGS2,KOS2,NK2loop}. The counterterms for ${\tilde S}_{ab \rightarrow 12}$ are the ultraviolet divergent coefficients times the basis color tensors. If there are $m$ color tensors, then the counterterms are  
\beq
S_L^{ab \rightarrow 12}= \sum_{I=1}^m c_I^{ab \rightarrow 12} Z_{IL}^{ab \rightarrow 12} \, ,
\label{ZLI}
\eeq
for the corrections to the color tensor $c_L^{ab \rightarrow 12}$, where $Z_{IL}^{ab \rightarrow 12}$ denotes the $IL$ matrix element of the renormalization matrix in Eq. (\ref{Zab}). 

Then
\beq
\Gamma_{\!\! S \, ab \rightarrow 12}=\left(\frac{dZ_{ab \rightarrow 12}}{d\ln\mu_R}\right) Z_{ab \rightarrow 12}^{-1}
=\beta(g_s, \epsilon)  \frac{\partial Z_{ab \rightarrow 12}}{\partial g_s} Z_{ab \rightarrow 12}^{-1} \, ,
\eeq
is the soft anomalous dimension matrix 
that controls the evolution of the soft function $S_{ab \rightarrow 12}$ via Eq. (\ref{rges}). 
In dimensional regularization $Z_{ab \rightarrow 12}$ has $1/\epsilon$ poles. 
Expanding $Z_{ab \rightarrow 12}$ in powers of the strong coupling, its $IL$ matrix element is
\beq 
Z_{IL \, ab \rightarrow 12}=\delta_{IL}+\frac{\alpha_s}{\pi}Z^{(1)}_{IL \, ab \rightarrow 12}+{\cal O}(\alpha_s^2) \, ,
\eeq 
and since $Z_{ab \rightarrow 12}^{(1)}$ has a $1/\epsilon$ pole while $\beta(g_s, \epsilon)$  
includes a $-g_s \epsilon/2$ term in dimensional regularization, 
we find that $\Gamma_{S \, ab \rightarrow 12}$ is given at one loop simply by minus the residue of $Z_{ab \rightarrow 12}$.

\subsection{Resummed cross section}

The resummed differential cross section in transform space is derived from the renormalization-group evolution of the $N$-dependent functions in Eq. (\ref{sigN}), i.e. ${\tilde \psi}/{\tilde \phi}$, ${\tilde J}$, and of course ${\tilde S}_{ab\rightarrow 12}$. We find \cite{NKGS1,NKGS2,LOS,NKtoprev}
\beqa
d{\hat{\sigma}}^{\rm resum}_{ab \rightarrow 12}(N) &=&   
\exp\left[\sum_{i=a,b} E_i(N_i)\right] 
\exp \left[\sum_{i=a,b} 2\int_{\mu_F}^{\sqrt{s}} \frac{d\mu}{\mu} \;
\gamma_{i/i}\left({\tilde N}_i,\alpha_s(\mu)\right)\right] 
\exp\left[\sum_{j={\rm f.s.} \, q,g} {E'}_j(N)\right] 
\nonumber\\ && \times
{\rm tr} \left \{H_{ab \rightarrow 12}\left(\alpha_s(\sqrt{s})\right) 
{\bar P} \exp \left[\int_{\sqrt{s}}^{{\sqrt{s}}/{\tilde N}} 
\frac{d\mu}{\mu} 
\Gamma_{\!\! S \, ab \rightarrow 12}^{\dagger}\left(\alpha_s(\mu)\right)\right] \right.
\nonumber\\ && \quad \quad \quad \left. \times
{\tilde S}_{ab \rightarrow 12} \left(\alpha_s(\frac{\sqrt{s}}{\tilde N})\right) 
P \exp \left[\int_{\sqrt{s}}^{{\sqrt{s}}/{\tilde N}} 
\frac{d\mu}{\mu} \Gamma_{\!\! S \, ab \rightarrow 12}
\left(\alpha_s(\mu)\right)\right] \right\} \, , 
\label{resHS}
\eeqa
where the symbols $P$ (${\bar P}$) denote path-ordering in the same (reverse) sense as the integration variable $\mu$, and ${\tilde N}=N e^{\gamma_E}$ with $\gamma_E$ the Euler constant.

The first exponential in Eq. (\ref{resHS}) resums soft and collinear contributions from the incoming partons \cite{GS87,CT89}. We have
\beq
E_i(N_i)=
\int^1_0 dz \frac{z^{N_i-1}-1}{1-z}\;
\left[\int_1^{(1-z)^2} \frac{d\lambda}{\lambda}
A_i\left(\alpha_s(\lambda s)\right)
+D_i\left(\alpha_s((1-z)^2 s)\right)\right] \, ,
\label{Eexp}
\eeq
with $A_i = \sum_{n=1}^{\infty}(\alpha_s/\pi)^n A_i^{(n)}$.
Here 
$A_i^{(1)}=C_i$ which is $C_F$ for a quark 
or antiquark and $C_A$ for a gluon, 
$A_i^{(2)}=C_i K/2$ where 
$K= C_A\; ( 67/18-\zeta_2) - 5n_f/9$ \cite{KT82} with $\zeta_2=\pi^2/6$,
$A_i^{(3)}$ is given by \cite{MVV04}
\beqa
A_i^{(3)}&=&C_i\left[C_A^2\left(\frac{245}{96}-\frac{67}{36}\zeta_2
+\frac{11}{24}\zeta_3+\frac{11}{8}\zeta_4\right)
+C_F n_f\left(-\frac{55}{96}+\frac{\zeta_3}{2}\right) \right.
\nonumber \\ && \quad \quad \left.
{}+C_A n_f \left(-\frac{209}{432}+\frac{5}{18}\zeta_2
-\frac{7}{12}\zeta_3\right)-\frac{n_f^2}{108}\right] \, ,
\eeqa
where $\zeta_4=\pi^4/90$, 
and $A_i^{(4)}$ is given by \cite{HKM,MPS}
\beqa
A_i^{(4)} &\!\!=\!\!& C_i \left[C_A^3 \left(\frac{42139}{10368}-\frac{5525}{1296}\zeta_2
+\frac{1309}{432}\zeta_3+\frac{451}{64}\zeta_4
-\frac{11}{24}\zeta_2\zeta_3-\frac{451}{288}\zeta_5
-\frac{\zeta_3^2}{16}-\frac{313}{96}\zeta_6\right) \right.
\nonumber \\ && \hspace{-5mm} 
{}+C_F^2 n_f \left(\frac{143}{576}+\frac{37}{48} \zeta_3
-\frac{5}{4} \zeta_5\right)
+C_F C_A n_f \left(-\frac{17033}{10368}+\frac{55}{96}\zeta_2
+\frac{29}{18}\zeta_3-\frac{11}{16}\zeta_4-\frac{1}{2}\zeta_2 \zeta_3
+\frac{5}{8}\zeta_5 \right)
\nonumber \\ && \hspace{-5mm} 
{}+C_A^2 n_f \left(-\frac{24137}{20736}
+\frac{635}{648} \zeta_2-\frac{361}{108}\zeta_3
-\frac{11}{48}\zeta_4+\frac{7}{12}\zeta_2\zeta_3
+\frac{131}{144}\zeta_5\right)
+C_F n_f^2\left(\frac{299}{2592}-\frac{5}{18} \zeta_3
+\frac{\zeta_4}{8} \right)
\nonumber \\ && \hspace{-5mm} \left. 
+C_A n_f^2 \left(\frac{923}{20736}-\frac{19}{648}\zeta_2
+\frac{35}{108}\zeta_3-\frac{7}{48}\zeta_4\right)
+n_f^3 \left(-\frac{1}{648}+\frac{\zeta_3}{108} \right) \right]
\nonumber \\ && \hspace{-5mm} 
{}+\frac{d_i^{abcd} d_A^{abcd}}{N_{R_i}} \left(-\frac{\zeta_2}{2}+\frac{\zeta_3}{6} 
+\frac{55}{12}\zeta_5-\frac{3}{2}\zeta_3^2-\frac{31}{8}\zeta_6\right)
+\frac{d_i^{abcd} d_F^{abcd}}{N_{R_i}} n_f \left(\zeta_2 
-\frac{\zeta_3}{3}-\frac{5}{3}\zeta_5\right)
\eeqa
where $\zeta_5=1.036927755\cdots$, $\zeta_6=\pi^6/945$, 
and where $d_i^{abcd}$ is $d_F^{abcd}$ for a quark or antiquark and $d_A^{abcd}$ for a gluon, 
while $N_{R_i}$ is $N_F=N_c$ for a quark or antiquark and $N_A=N_c^2-1$ for a gluon.

Also $D_i=\sum_{n=1}^{\infty}(\alpha_s/\pi)^n D_i^{(n)}$, 
with $D_i^{(1)}=0$, $D_i^{(2)}$ given by \cite{CLS97}
\beq
D_i^{(2)}=C_i \left[C_A \left(-\frac{101}{54}+\frac{11}{6} \zeta_2
+\frac{7}{4}\zeta_3\right)
+n_f \left(\frac{7}{27}-\frac{\zeta_2}{3}\right)\right] \, ,
\eeq
and $D_i^{(3)}$ given by \cite{MA05} 
\beqa
D_i^{(3)}\!\!\!\!&=\!\!\!&C_i \left[C_A^2 \left(-\frac{297029}{46656}
+\frac{6139}{648} \zeta_2+\frac{2509}{216} \zeta_3
-\frac{187}{48} \zeta_4 -\frac{11}{12} \zeta_2 \zeta_3-3 \zeta_5\right)
+n_f^2 \left(-\frac{29}{729}+\frac{5}{27}\zeta_2
+\frac{5}{54}\zeta_3 \right) \right.
\nonumber \\ && \quad \quad\left.
{}+C_A n_f \left(\frac{31313}{23328}-\frac{1837}{648}\zeta_2
-\frac{155}{72}\zeta_3+\frac{23}{24}\zeta_4\right)
+C_F n_f \left(\frac{1711}{1728}-\frac{\zeta_2}{4}-\frac{19}{36}\zeta_3
-\frac{\zeta_4}{4}\right)\right] \, .
\eeqa

We note that our expression for the resummed cross section is for hadron-hadron scattering but can be easily adapted for hadron-lepton scattering (the sums over $a$, $b$ in the first two exponentials reduce to one term) and lepton-lepton scattering (the sums vanish).

In the second exponential in Eq. (\ref{resHS}), involving the factorization scale, 
$\gamma_{i/i}$ is the moment-space 
anomalous dimension of the ${\overline {\rm MS}}$ density $\phi_{i/i}$ \cite{FRS1,FRS2,GALY79,GFP80,FP80},
$\gamma_{i/i}=-A_i \ln {\tilde N_i} +\gamma_i$ with parton anomalous dimensions
$\gamma_i=\sum_{n=1}^{\infty}(\alpha_s/\pi)^n \gamma_i^{(n)}$,
where $\gamma_q^{(1)}=3C_F/4$, $\gamma_g^{(1)}=\beta_0/4$,
\beq
\gamma_q^{(2)}=C_F^2\left(\frac{3}{32}-\frac{3}{4}\zeta_2
+\frac{3}{2}\zeta_3\right)
+C_F C_A\left(\frac{17}{96}+\frac{11}{12}\zeta_2-\frac{3}{4}\zeta_3\right)
-C_F n_f\left(\frac{1}{48}+\frac{\zeta_2}{6} \right)\, ,
\eeq
and
\beq
\gamma_g^{(2)}=C_A^2\left(\frac{2}{3}+\frac{3}{4}\zeta_3\right)
-n_f\left(\frac{C_F}{8}+\frac{C_A}{6}\right) \, .
\eeq
Also $\gamma_q^{(3)}$ is given by \cite{MVV04}
\beqa
\gamma_q^{(3)}&=& C_F^3 \left(\frac{29}{128}+\frac{9}{32} \zeta_2+\frac{17}{16}\zeta_3
+\frac{9}{4}\zeta_4-\frac{1}{2}\zeta_2\zeta_3-\frac{15}{4}\zeta_5 \right)
+C_F^2 n_f \left(-\frac{23}{64} +\frac{5}{48}\zeta_2 
-\frac{17}{24}\zeta_3 + \frac{29}{48}\zeta_4 \right)
\nonumber \\ && \hspace{-11mm}
{}+C_F n_f^2 \left(-\frac{17}{576}+\frac{5}{108} \zeta_2 
-\frac{\zeta_3}{36} \right)
+C_F^2 C_A\left(\frac{151}{256}-\frac{205}{96} \zeta_2+\frac{211}{48}\zeta_3
-\frac{247}{96}\zeta_4+\frac{1}{4}\zeta_2 \zeta_3+\frac{15}{8} \zeta_5 \right)
\nonumber \\ && \hspace{-11mm}
{}+C_F C_A^2 \left(-\frac{1657}{2304} 
+\frac{281}{108}\zeta_2-\frac{97}{36}\zeta_3 
-\frac{5}{64}\zeta_4+\frac{5}{8} \zeta_5 \right)
+C_F C_A n_f  \left(\frac{5}{16} 
-\frac{167}{216} \zeta_2+\frac{25}{72} \zeta_3 
+\frac{\zeta_4}{32} \right)\!.
\eeqa

The third exponential in Eq. (\ref{resHS}) describes soft and collinear radiation from the final-state (f.s.) particles \cite{GS87,CT89,KOS1,LOS}. The exponent vanishes for colorless particles and for massive particles, i.e. ${E'}_j=0$ in those cases. In the case of hadron production, where a quark or gluon hadronizes into the observed hadron, then this exponent is the same as the initial-state exponent \cite{HRSV}, i.e. ${E'}_1(N)=E_1(N)$ plus the same term in the exponent for the scale dependence. For heavy jets there is a different expression \cite{KOS1}. For all other cases with final-state massless quarks or gluons, which is the majority of the cases studied, the exponent has the expression 
\beq
{E'}_j(N)=
\int^1_0 dz \frac{z^{N-1}-1}{1-z}\;
\left[\int^{1-z}_{(1-z)^2} \frac{d\lambda}{\lambda}
A_j \left(\alpha_s\left(\lambda s\right)\right)
+B_j\left(\alpha_s((1-z)s)\right)
+D_j\left(\alpha_s((1-z)^2 s)\right)\right] \, ,
\label{Ejexp}
\eeq
where $B_j=\sum_{n=1}^{\infty}(\alpha_s/\pi)^n B_j^{(n)}$
with $B_q^{(1)}=-3C_F/4$ and $B_g^{(1)}=-\beta_0/4$.
Also for $B_q^{(2)}$ \cite{CLS97} and $B_g^{(2)}$ (cf. \cite{MVV05}), we have 
\beq
B_q^{(2)}=C_F^2\left(-\frac{3}{32}+\frac{3}{4}\zeta_2-\frac{3}{2}\zeta_3\right)
+C_F C_A \left(-\frac{57}{32}-\frac{11}{12}\zeta_2+\frac{3}{4}\zeta_3\right)
+C_F n_f \left(\frac{5}{16}+\frac{\zeta_2}{6}\right) \, ,
\eeq
and 
\beq
B_g^{(2)}=C_A^2\left(-\frac{1025}{432}-\frac{3}{4}\zeta_3\right)
+\frac{79}{108} C_A \, n_f +C_F \frac{n_f}{8}-\frac{5}{108} n_f^2 \, ,
\eeq
while $B_q^{(3)}$ and $B_g^{(3)}$ are given by (cf. \cite{MVV05})
\beqa
B_q^{(3)}&=&C_F^3 \left(-\frac{29}{128}-\frac{9}{32} \zeta_2-\frac{17}{16}\zeta_3 
-\frac{9}{4}\zeta_4+\frac{\zeta_2\zeta_3}{2}+\frac{15}{4}\zeta_5 \right)
+C_F^2 n_f \left(\frac{77}{128} - \frac{17}{32}\zeta_2 
+\frac{7}{12}\zeta_3 + \frac{\zeta_4}{4} \right)
\nonumber \\ && \hspace{-10mm}
{}+C_F C_A n_f  \left(\frac{455}{216} 
+\frac{199}{216} \zeta_2+\frac{29}{36} \zeta_3 
-\frac{5}{12}\zeta_4 \right)
+C_F C_A^2 \left(-\frac{5599}{864} 
-\frac{2831}{864}\zeta_2+\frac{5}{9}\zeta_3 
+\frac{211}{96}\zeta_4
-\frac{5}{8} \zeta_5 \right)
\nonumber \\ && \hspace{-10mm}
{}+C_F^2 C_A\left(-\frac{23}{32}+\frac{287}{64} \zeta_2-\frac{89}{24}\zeta_3
-\frac{17}{8}\zeta_4 -\frac{\zeta_2 \zeta_3}{4}-\frac{15}{8} \zeta_5 \right)
+C_F n_f^2 \left(-\frac{127}{864}-\frac{11}{216} \zeta_2 
-\frac{\zeta_3}{9} \right)
\eeqa
and 
\beqa
B_g^{(3)}&=&C_A^3 \left(-\frac{299341}{31104}
+\frac{1307}{576} \zeta_2
-\frac{523}{144} \zeta_3-\frac{275}{96} \zeta_4 
+\frac{\zeta_2 \zeta_3}{4} + \frac{5}{4} \zeta_5 \right) 
\nonumber \\ && \hspace{-4mm} 
{}+C_A^2 n_f \left(\frac{41453}{10368}-\frac{39}{32} \zeta_2
+\frac{59}{36} \zeta_3+\frac{25}{48}\zeta_4 \right)
+C_A C_F n_f \left(\frac{191}{128}-\frac{11}{12} \zeta_3\right)
-C_F^2\frac{n_f}{64}
\nonumber \\ && \hspace{-4mm}
{}+C_A n_f^2 \left(-\frac{557}{1152}+\frac{11}{48} \zeta_2
-\frac{7}{36} \zeta_3 \right)
+C_F n_f^2 \left(-\frac{47}{192}+\frac{\zeta_3}{6} \right) 
+n_f^3 \left(\frac{25}{1944}-\frac{\zeta_2}{72} \right).
\eeqa

For the hard and soft functions we use the expansions 
$H_{ab \rightarrow 12}=\sum_{n=0}^{\infty}(\alpha_s^{d+n}/\pi^n) \, H_{ab \rightarrow 12}^{(n)}$, where $d$ is the power of $\alpha_s$ in the leading-order cross section and it depends on the process, 
and
$S_{ab \rightarrow 12}=\sum_{n=0}^{\infty}(\alpha_s/\pi)^n \, S_{ab \rightarrow 12}^{(n)}$.
At lowest order, the soft matrix is given in terms of the color tensor basis by the expression $S^{(0)}_{LI\, ab \rightarrow 12}={\rm tr}\left(c_L^{* \, ab\rightarrow 12} c_I^{ab \rightarrow 12}\right)$, while the hard matrix is real and symmetric, $H^{(0)}_{LI\, ab \rightarrow 12}=h_{L \, ab \rightarrow 12}^{(0)} h_{I \, ab \rightarrow 12}^{(0) *}$ with $h_{L \, ab \rightarrow 12}^{(0)}=(S^{(0)}_{LK \, ab \rightarrow 12})^{-1} {\rm tr}(c_K^{ab \rightarrow 12 \, *} M_{ab \rightarrow 12}^{(0)})$ and $h_{I \, ab \rightarrow 12}^{(0) *}={\rm tr}(M_{ab \rightarrow 12}^{(0) \, \dagger} c_K^{ab \rightarrow 12}) (S^{(0)}_{KI \, ab \rightarrow 12})^{-1}$ where $M_{ab \rightarrow 12}^{(0)}$ is the lowest-order amplitude.

For the soft anomalous dimension we use the expansion 
$\Gamma_{\!\! S \, ab \rightarrow 12}=\sum_{n=1}^{\infty} (\alpha_s/\pi)^n \, \Gamma_{\!\! S \, ab \rightarrow 12}^{(n)}$.
In the past, through the year 2008, most expressions for soft anomalous dimensions in the literature were given at one loop in axial gauge, while since 2009 most have been given in Feynman gauge. The relation between the two is
$\Gamma_{\!\! S \, ab \rightarrow 12}^{(1)\, {\rm axial}}=\Gamma_{\!\! S \, ab \rightarrow 12}^{(1)\, {\rm Feyn.}}+(1/2)[A_a^{(1)}+A_b^{(1)}+\sum_{j={\rm f.s.} \, q,g} A_j^{(1)}]$. Of course, the overall result for the resummed cross section is gauge-independent, and this is easily seen at one-loop from the fact that in axial gauge $D_i^{(1)\, {\rm axial}}=-A_i^{(1)}$, which precisely cancel the extra terms from change of gauge in the soft anomalous dimension.

We also note that in general we can disregard imaginary $i \pi$ terms in the soft anomalous dimensions since such terms generally do not contribute. Of course, in cases with simple color flow when we only have functions - not matrices - it is clear from Eq. (\ref{resHS}) that any imaginary terms in $\Gamma_{\!\! S \, ab \rightarrow 12}$ cancel out against imaginary terms in its Hermitian adjoint in the resummed cross section. Such terms also routinely cancel and do not contribute in fixed-order expansions even when we have processes that require matrices. Therefore, in the results in Sections 5 through 10 we will drop any $i\pi$ terms for simplicity.  

In addition, we note that one can perform resummation in other schemes such as the DIS scheme (see e.g. Refs. \cite{NKGS1,NKGS2,NKun04} for the relation between the resummed expressions for the cross section in $\rm {\overline {MS}}$ and DIS schemes). The expressions for the soft anomalous dimensions, however, remain the same. 

When all external eikonal lines in the scattering process are massless, then the two-loop soft anomalous dimension matrix is proportional to the one-loop result \cite{ADS}. Furthermore, three-parton correlations with massless eikonal lines do not contribute to the soft anomalous dimension at any order due to scaling symmetry constraints \cite{BN,GM}. However, at three loops for the soft anomalous dimension in massless multi-leg scattering there are contributions from four-parton correlations \cite{ADG}. 

When there are massive eikonal lines then the soft anomalous dimension at two or more loops is no longer proportional to the one-loop result \cite{NK2loop}. Also, when two of the eikonal lines in the scattering process are massive, then the three-parton correlations no longer vanish; however, three-parton correlations still vanish when only one of the eikonal lines is massive.

\mysection{NLO, NNLO, and N$^3$LO expansions of the resummed cross section}

In this section we expand the resummed cross section to NLO, NNLO, and N$^3$LO, using Eqs. ({\ref{resHS}), (\ref{Eexp}), and (\ref{Ejexp}) (see also \cite{NKun04,NKNNNLO}). Our expansions can be used for a variety of processes, but with the restrictions noted for the use of Eq. (\ref{Ejexp}) in the previous section.

The corrections take the form of plus distributions
\beq
{\cal D}_k(s_4)=\left[\frac{\ln^k(s_4/s)}{s_4}\right]_+ \, , 
\eeq
defined by
\beq
\int_0^{s_{4 \, {\rm max}}} ds_4 \, f(s_4) \left[\frac{\ln^k(s_4/s)}{s_4}\right]_{+}=\int_0^{s_{4\, {\rm max}}} ds_4 \frac{\ln^k(s_4/s)}{s_4} \left(f(s_4) - f(0)\right)
+\frac{1}{k+1} \ln^{k+1}\left(\frac{s_{4\, {\rm max}}}{s}\right) f(0) \, , 
\eeq
where $f$ is a smooth function, such as a pdf.
Of course the above distributions, involving logarithms of $s_4/s$, can readily be reexpressed in terms of logarithms of $s_4/M^2$ for any hard scale $M$ relevant to the process considered \cite{NKun04,NKNNNLO}. 

We note that the expressions can be simplified \cite{NKun04} for the cases of simple color structure where the soft anomalous dimensions are not matrices, as we will discuss in more detail below. 
We also note that we can extend the formulas for the fixed-order expansions to the general case of multi-particle final states, i.e. $2 \to n$ processes \cite{MFNK}, by replacing $m_2^2$ by $(p_2+\cdots+p_n)^2$ in the formulas.
 
\subsection{NLO soft-gluon corrections}

The NLO soft-gluon corrections are
\beqa
d{\hat{\sigma}}_{ab \rightarrow 12}^{(1)} &=& F^{LO}_{ab \rightarrow 12} \, \frac{\alpha_s(\mu_R)}{\pi}
\left[c_3\, {\cal D}_1(s_4) + c_2\,  {\cal D}_0(s_4) 
+c_1\,  \delta(s_4)\right]
\nonumber \\ &&
{}+\frac{\alpha_s^{d+1}(\mu_R)}{\pi} 
\left[A_{ab \rightarrow 12} \, {\cal D}_0(s_4)+T_{ab \rightarrow 12} \, \delta(s_4)\right] \, ,
\label{NLOmaster}
\eeqa
where 
$F^{LO}_{ab \rightarrow 12}=\alpha_s^d \, {\rm tr} \left(H_{ab \rightarrow 12}^{(0)} S_{ab \rightarrow 12}^{(0)} \right)$ denotes the leading-order (LO) coefficient, 
\beq
c_3=2 (A_a^{(1)}+A_b^{(1)}) -\! \sum_{j={\rm f.s.} \, q,g} \!\!\! A_j^{(1)}\, ,
\label{c3}
\eeq 
and $c_2$ is given by $c_2=c_2^{\mu}+T_2$, 
with
\beq
c_2^{\mu}=-(A_a^{(1)}+A_b^{(1)}) \ln\left(\frac{\mu_F^2}{s}\right)
\eeq
denoting the terms involving logarithms of the scale, and  
\beq
T_2=-2 \, A_a^{(1)} \, \ln\left(\frac{m_2^2-u}{s}\right)
-2 \, A_b^{(1)} \, \ln\left(\frac{m_2^2-t}{s}\right) +D_a^{(1)}+D_b^{(1)}
+\! \sum_{j={\rm f.s.} \, q,g} \!\! \left(B_j^{(1)}+D_j^{(1)}\right) \, 
\label{T2}
\eeq
denoting the scale-independent terms.
Also,
\beq
A_{ab \rightarrow 12}={\rm tr} \left(H_{ab \rightarrow 12}^{(0)} \, \Gamma_{\!\! S \, ab \rightarrow 12}^{(1)\,\dagger} \, S_{ab \rightarrow 12}^{(0)}
+H_{ab \rightarrow 12}^{(0)} \, S_{ab \rightarrow 12}^{(0)} \, \Gamma_{\!\! S \, ab \rightarrow 12}^{(1)}\right) \, .
\label{Ac}
\eeq
For the cases with simple color structure where the soft anomalous dimension is not a matrix, we have $\alpha_s^d A_{ab \rightarrow 12}= 2 \, {\rm Re}\Gamma_{\!\! S \, ab \rightarrow 12}^{(1)} F^{\rm LO}_{ab \rightarrow 12}$, and thus this term can be added to the term $c_2$, thus simplifying the expressions at NLO and higher orders \cite{NKun04}.

The $\delta(s_4)$ terms involve a term
$c_1$, that is proportional 
to the Born cross section, and a term $T_{ab \rightarrow 12}$ that, in general, is not.
We write $c_1 =c_1^{\mu} +T_1$, with
\beq
c_1^{\mu}=\left[A_a^{(1)}\, \ln\left(\frac{m_2^2-u}{s}\right) 
+A_b^{(1)}\, \ln\left(\frac{m_2^2-t}{s}\right)
-\gamma_a^{(1)}-\gamma_b^{(1)}\right]\ln\left(\frac{\mu_F^2}{s}\right)
+d \frac{\beta_0}{4} \ln\left(\frac{\mu_R^2}{s}\right) 
\label{c1mu}
\eeq
denoting the terms involving logarithms of the scale.
Also 
\beq
T_1=A_a^{(1)}\, \ln^2\left(\frac{m_2^2-u}{s}\right)
+A_b^{(1)}\, \ln^2\left(\frac{m_2^2-t}{s}\right)
-D_a^{(1)} \, \ln\left(\frac{m_2^2-u}{s}\right)
-D_b^{(1)} \, \ln\left(\frac{m_2^2-t}{s}\right)
\eeq
and 
\beq
T_{ab \rightarrow 12}={\rm tr} \left(H_{ab \rightarrow 12}^{(0)} \,  S_{ab \rightarrow 12}^{(1)}+H_{ab \rightarrow 12}^{(1)} \, S_{ab \rightarrow 12}^{(0)} \right) \, .
\eeq
We note that $T_{ab \rightarrow 12}$ can also be determined via a comparison to a complete NLO calculation.

\subsection{NNLO soft-gluon corrections}

The NNLO soft-gluon corrections are 
\beqa
d{\hat{\sigma}}_{ab \rightarrow 12}^{(2)}&=&F^{LO}_{ab \rightarrow 12} \frac{\alpha_s^2(\mu_R)}{\pi^2} \left\{\frac{1}{2}c_3^2\, {\cal D}_3(s_4) 
+\left[\frac{3}{2}c_3 c_2-\frac{\beta_0}{4} c_3
+\frac{\beta_0}{8} \! \sum_{j={\rm f.s.} \, q,g} \!\!\! A_j^{(1)}\right]  {\cal D}_2(s_4) \right.
\nonumber \\ && \hspace{-21mm}
{}+\left[c_3 c_1+c_2^2-\zeta_2 c_3^2
-\frac{\beta_0}{2} T_2+\frac{\beta_0}{4} c_3 
\ln\left(\frac{\mu_R^2}{s}\right)+2(A_a^{(2)}+A_b^{(2)})
+\! \sum_{j={\rm f.s.} \, q,g} \!\! \left(-A_j^{(2)}+\frac{\beta_0}{4} B_j^{(1)}\right)\right] {\cal D}_1(s_4)
\nonumber \\ && \hspace{-21mm} 
{}+\left[c_2 c_1-\zeta_2 c_3 c_2+\zeta_3 c_3^2-\frac{\beta_0}{2}T_1
+\frac{\beta_0}{4} c_2 \ln\left(\frac{\mu_R^2}{s}\right) 
-2 A_a^{(2)} \ln\left(\frac{m_2^2-u}{s}\right)
-2 A_b^{(2)}\ln\left(\frac{m_2^2-t}{s}\right) \right.
\nonumber \\ && \hspace{-13mm} 
{}+D_a^{(2)}+D_b^{(2)}+\frac{\beta_0}{8} (A_a^{(1)}+A_b^{(1)}) \ln^2\left(\frac{\mu_F^2}{s}\right)-(A_a^{(2)}+A_b^{(2)}) \ln\left(\frac{\mu_F^2}{s}\right)
\nonumber \\ && \hspace{-13mm} \left. \left.
{}+\! \sum_{j={\rm f.s.} \, q,g} \!\! \left(B_j^{(2)}+D_j^{(2)}\right) \right] 
 {\cal D}_0(s_4) \right\}
\nonumber \\ && \hspace{-25mm}
{}+\frac{\alpha_s^{d+2}(\mu_R)}{\pi^2} 
\left\{\frac{3}{2} c_3 A_{ab \rightarrow 12} \, {\cal D}_2(s_4)
+\left[\left(2 c_2-\frac{\beta_0}{2}\right) A_{ab \rightarrow 12}
+c_3 T_{ab \rightarrow 12} +F_{ab \rightarrow 12}\right]
{\cal D}_1(s_4) \right. 
\nonumber \\ && \left. 
{}+\left[\left(c_1-\zeta_2 c_3+\frac{\beta_0}{4}\ln\left(\frac{\mu_R^2}{s}
\right)\right)A_{ab \rightarrow 12}+c_2 T_{ab \rightarrow 12}+G_{ab \rightarrow 12}\right]
{\cal D}_0(s_4) \right\} \, ,
\label{NNLOmaster}
\eeqa
where
\beqa
F_{ab \rightarrow 12}&=&{\rm tr} \left[H_{ab \rightarrow 12}^{(0)} \, \left(\Gamma_{\!\! S \, ab \rightarrow 12}^{(1)\,\dagger}\right)^2 \, S_{ab \rightarrow 12}^{(0)}
+H_{ab \rightarrow 12}^{(0)} \, S_{ab \rightarrow 12}^{(0)} \, \left(\Gamma_{\!\! S \, ab \rightarrow 12}^{(1)}\right)^2 \right.
\nonumber \\ && \quad \left. 
{}+2 \, H_{ab \rightarrow 12}^{(0)} \, \Gamma_{\!\! S \, ab \rightarrow 12}^{(1)\,\dagger} \, S_{ab \rightarrow 12}^{(0)} \, \Gamma_{\!\! S \, ab \rightarrow 12}^{(1)} \right] 
\label{Fterm}
\eeqa
and
\beqa
G_{ab \rightarrow 12}&=&{\rm tr} \left[H_{ab \rightarrow 12}^{(1)} \, \Gamma_{\!\! S \, ab \rightarrow 12}^{(1)\,\dagger} \, S_{ab \rightarrow 12}^{(0)}
+H_{ab \rightarrow 12}^{(1)} \, S_{ab \rightarrow 12}^{(0)} \, \Gamma_{\!\! S \, ab \rightarrow 12}^{(1)} \right.
\nonumber \\ && \quad
{}+H_{ab \rightarrow 12}^{(0)} \, \Gamma_{\!\! S \, ab \rightarrow 12}^{(1)\,\dagger} \, S_{ab \rightarrow 12}^{(1)} 
+H_{ab \rightarrow 12}^{(0)} \, S_{ab \rightarrow 12}^{(1)} \, \Gamma_{\!\! S \, ab \rightarrow 12}^{(1)} 
\nonumber \\ && \quad \left. 
{}+H_{ab \rightarrow 12}^{(0)} \, \Gamma_{\!\! S \, ab \rightarrow 12}^{(2)\,\dagger} \, S_{ab \rightarrow 12}^{(0)}
+H_{ab \rightarrow 12}^{(0)} \, S_{ab \rightarrow 12}^{(0)} \, \Gamma_{\!\! S \, ab \rightarrow 12}^{(2)} \right] \, .
\eeqa

\subsection{N$^3$LO soft-gluon corrections}

The N$^3$LO soft-gluon corrections are
\beqa
d{\hat{\sigma}}_{ab \rightarrow 12}^{(3)}&=& 
F^{LO}_{ab \rightarrow 12} \frac{\alpha_s^3(\mu_R)}{\pi^3} \left\{  
\frac{1}{8} \, c_3^3 \, {\cal D}_5(s_4)
+\left[\frac{5}{8} \, c_3^2 \, c_2 -\frac{5}{24} \, c_3^2 \, \beta_0
+\frac{5}{48} c_3 \beta_0 \! \sum_{j={\rm f.s.} \, q,g} \!\!\! A_j^{(1)} \right]  
{\cal D}_4(s_4) \right.
\nonumber \\ && \hspace{-8mm}
{}+\left[c_3 c_2^2 +\frac{1}{2}c_3^2 c_1-\zeta_2 c_3^3 
+\frac{\beta_0^2}{12}c_3-\frac{5}{6}\beta_0 c_3 c_2
+\frac{\beta_0}{4}c_3^2 \ln\left(\frac{\mu_R^2}{s}\right)
-\frac{\beta_0}{2}c_3 \left(A_a^{(1)}+A_b^{(1)}\right) 
\ln\left(\frac{\mu_F^2}{s}\right) \right.
\nonumber \\ &&   \left. 
{}+2 c_3 \left(A_a^{(2)}+A_b^{(2)}\right)
+\! \sum_{j={\rm f.s.} \, q,g} \!\! \left(-\frac{\beta_0^2}{16}A_j^{(1)} -c_3 A_j^{(2)}
+\frac{\beta_0}{6} c_2 A_j^{(1)}+c_3 \frac{\beta_0}{4} B_j^{(1)} \right)
\right] {\cal D}_3(s_4)
\nonumber \\ && \hspace{-8mm}
{}+\left[\frac{3}{2}\, c_3 \,c_2 \, c_1 +\frac{1}{2} \, c_2^3
-3\, \zeta_2 \, c_3^2 \,c_2 +\frac{5}{2} \, \zeta_3 \, c_3^3
-\frac{\beta_0}{4} c_3 c_1 +\frac{3}{4} \beta_0 \zeta_2 c_3^2 
-\frac{\beta_0}{4}\left(3 c_2-\beta_0 \right) T_2 \right.
\nonumber \\ &&  
{}+\frac{\beta_0}{8} c_3 \left(6 c_2 -\beta_0 \right) 
\ln\left(\frac{\mu_R^2}{s}\right)
+(3 c_2-\beta_0) \left(A_a^{(2)}+A_b^{(2)}\right)
-\left(A_a^{(1)}+A_b^{(1)}\right) \frac{\beta_1}{8}
\nonumber \\ && 
{}-\frac{3}{2}c_3\left[\frac{\beta_0}{2} T_1
+2 A_a^{(2)} \ln\left(\frac{m_2^2-u}{s}\right)
+2 A_b^{(2)} \ln\left(\frac{m_2^2-t}{s}\right) \right.
\nonumber \\ && \quad \quad \quad \left.
{}-D_a^{(2)}-D_b^{(2)}
-\frac{\beta_0}{8} \left(A_a^{(1)}+A_b^{(1)}\right) 
\ln^2\left(\frac{\mu_F^2}{s}\right)
+\left(A_a^{(2)}+A_b^{(2)}\right) \ln\left(\frac{\mu_F^2}{s}\right)\right]
\nonumber \\ && 
{}-\frac{3}{2} c_3 \! \sum_{j={\rm f.s.} \, q,g} \!\! \left(-B_j^{(2)}-D_j^{(2)}
+\frac{\beta_0}{4} \zeta_2 A_j^{(1)}\right)
+3 c_2 \! \sum_{j={\rm f.s.} \, q,g} \!\! \left(-\frac{A_j^{(2)}}{2}
+\frac{\beta_0}{8} B_j^{(1)}\right)
\nonumber \\ &&   \left.
{}+\! \sum_{j={\rm f.s.} \, q,g} \!\! 
\left(A_j^{(1)} \frac{\beta_0^2}{16} \ln\left(\frac{\mu_R^2}{s}\right)
+\frac{3\beta_0}{4} A_j^{(2)}-\frac{3\beta_0^2}{16} B_j^{(1)}
+\frac{3}{32} A_j^{(1)} \beta_1+\frac{\beta_0}{8} c_1 A_j^{(1)}\right) 
\right] {\cal D}_2(s_4)
\nonumber \\ && \hspace{-8mm}  \left.
{}+{\cal O}\left(D_1(s_4)\right) \right\} 
\nonumber \\ && \hspace{-13mm}
{}+\frac{\alpha_s^{d+3}(\mu_R)}{\pi^3} \left\{
\frac{5}{8} \, c_3^2 \, A_{ab \rightarrow 12} \, {\cal D}_4(s_4) \right.
\nonumber \\ &&
+\left[\frac{1}{2}\, c_3^2\, T_{ab \rightarrow 12}
+\left(2\, c_3 \, c_2 - \frac{5}{6}\beta_0 \, c_3  
+\frac{\beta_0}{6} \! \sum_{j={\rm f.s.} \, q,g} \!\!\! A_j^{(1)}\right) 
\, A_{ab \rightarrow 12} +c_3 \, F_{ab \rightarrow 12}\right] {\cal D}_3(s_4) 
\nonumber \\ && \hspace{-8mm}
{}+\left[\left(\frac{3}{2}\, c_3 \, c_2-\frac{\beta_0}{4} c_3
+\frac{\beta_0}{8}\! \sum_{j={\rm f.s.} \, q,g} \!\!\! A_j^{(1)} \right) \, 
T_{ab \rightarrow 12} 
+\frac{3}{2}\, c_2\, F_{ab \rightarrow 12} +\frac{3}{2} \, c_3 \, G_{ab \rightarrow 12} +\frac{1}{2} \, K_{ab \rightarrow 12} \right.
\nonumber \\ &&
{}+\left(\frac{3}{2}\, c_3 \, c_1+\frac{3}{2}\, c_2^2
-3 \, \zeta_2 \, c_3^2 
-\frac{3}{4}\beta_0 (c_2+T_2)
+\frac{3}{4}\beta_0 c_3 \ln\left(\frac{\mu_R^2}{s}\right)
+\frac{\beta_0^2}{4} \right.
\nonumber \\ && \quad \quad \left. \left.
{}+3 A_a^{(2)}+3 A_b^{(2)}+\frac{3}{2} \! \sum_{j={\rm f.s.} \, q,g} \!\!
(-A_j^{(2)}+\frac{\beta_0}{4} B_j^{(1)}) \right) \, A_{ab \rightarrow 12} \right] {\cal D}_2(s_4) 
\nonumber \\ && \hspace{-8mm} \left.
{}+{\cal O}\left(D_1(s_4)\right) \right\} 
\label{NNNLOmaster}
\eeqa
where 
\beqa
\hspace{-15mm} K_{ab \rightarrow 12}\!\!&=&\!{\rm tr} 
\left[H_{ab \rightarrow 12}^{(0)} \, (\Gamma_{\!\! S \, ab \rightarrow 12}^{(1) \dagger})^3 \, S_{ab \rightarrow 12}^{(0)}
+H_{ab \rightarrow 12}^{(0)} \, S_{ab \rightarrow 12}^{(0)} \, (\Gamma_{\!\! S \, ab \rightarrow 12}^{(1)})^3 \right.
\nonumber \\ && \quad \left.
{}+3 \, H_{ab \rightarrow 12}^{(0)} \, (\Gamma_{\!\! S \, ab \rightarrow 12}^{(1)\dagger})^2 \, S_{ab \rightarrow 12}^{(0)} \, \Gamma_{\!\! S \, ab \rightarrow 12}^{(1)}
+ 3 \, H_{ab \rightarrow 12}^{(0)} \, \Gamma_{\!\! S \, ab \rightarrow 12}^{(1) \dagger} \, 
S_{ab \rightarrow 12}^{(0)} \, (\Gamma_{\!\! S \, ab \rightarrow 12}^{(1)})^2 \right] \, ,
\eeqa
and where we have omitted terms of order $D_1$ and $D_0$ (see also Ref. \cite{NKNNNLO}).

\mysection{Cusp anomalous dimension}

The cusp anomalous dimension \cite{NK2loop,AMP,BNS,IKR,KR,GHKM,NK3loopcusp}, involving two eikonal lines, is the simplest soft anomalous dimension, and it is an essential ingredient for all calculations of soft anomalous dimensions.

The cusp angle $\theta$ was introduced in Sec. 2.2, and it is given by 
\beq
\theta=\ln\left(\frac{p_i \cdot p_j+\sqrt{(p_i \cdot p_j)^2-p_i^2 p_j^2}}{\sqrt{p_i^2 \, p_j^2}}\right)=\ln\left(\frac{v_i \cdot v_j+\sqrt{(v_i \cdot v_j)^2-v_i^2 v_j^2}}{\sqrt{v_i^2 \, v_j^2}}\right)
\eeq
or equivalently $\theta=\cosh^{-1}(p_i\cdot p_j/\sqrt{p_i^2 p_j^2})=\cosh^{-1}(v_i\cdot v_j/\sqrt{v_i^2 v_j^2})$, where $p^{\mu}_i=(\sqrt{s}/2) v^{\mu}_i$ and $p^{\mu}_j=(\sqrt{s}/2) v^{\mu}_j$, where $v^{\mu}_i$ and $v^{\mu}_j$ are four-velocities.

The perturbative series for the cusp anomalous dimension in QCD is written as 
\beq
\Gamma_{\rm cusp}=\sum_{n=1}^{\infty} \left(\frac{\alpha_s}{\pi}\right)^n \Gamma^{(n)}_{\rm cusp}
\eeq 
where $\alpha_s$ is the strong coupling.

The cusp anomalous dimension at one loop \cite{AMP} is given by
\beq
\Gamma^{(1)}_{\rm cusp}=C_F (\theta \coth\theta -1) \, .
\label{cusp1loop}
\eeq 

The cusp anomalous dimension at two loops \cite{NK2loop,KR} is given by 
\beq
\Gamma^{(2)}_{\rm cusp}=K_2 \, \Gamma^{(1)}_{\rm cusp}+C^{(2)}
\eeq
where $C^{(2)}=C_F C_A C^{' (2)}$,
with \cite{NK2loop}
\beqa
C^{'(2)}&=&\frac{1}{2}+\frac{\zeta_2}{2}+\frac{\theta^2}{2} 
-\frac{1}{2}\coth\theta\left[\zeta_2\theta+\theta^2
+\frac{\theta^3}{3}+{\rm Li}_2\left(1-e^{-2\theta}\right)\right]
\nonumber \\ && \hspace{-3mm}
{}+\frac{1}{2}\coth^2\theta\left[-\zeta_3+\zeta_2\theta+\frac{\theta^3}{3}
+\theta \, {\rm Li}_2\left(e^{-2\theta}\right)
+{\rm Li}_3\left(e^{-2\theta}\right)\right] \, ,
\label{cusp2loop}
\eeqa
and where 
\beq
K_2=C_A \left(\frac{67}{36}-\frac{\zeta_2}{2}\right)-\frac{5}{18} n_f \, ,
\eeq
i.e. $K_2=K/2=A_i^{(2)}/C_i$.

The cusp anomalous dimension at three loops \cite{GHKM,NK3loopcusp} is given by 
\beq
\Gamma^{(3)}_{\rm cusp}= K_3 \, \Gamma^{(1)}_{\rm cusp}
+2 K_2 \left(\Gamma^{(2)}_{\rm cusp}-K_2 \, \Gamma^{(1)}_{\rm cusp}\right)+C^{(3)} 
= K_3 \, \Gamma^{(1)}_{\rm cusp}
+2 K_2 \, C^{(2)}+C^{(3)}  \, ,
\label{cusp3loop}
\eeq
where
\beq
K_3=C_A^2\left(\frac{245}{96}-\frac{67}{36}\zeta_2
+\frac{11}{24}\zeta_3+\frac{11}{8}\zeta_4\right)
+C_F n_f\left(-\frac{55}{96}+\frac{\zeta_3}{2}\right)
+C_A n_f \left(-\frac{209}{432}+\frac{5\zeta_2}{18}
-\frac{7\zeta_3}{12}\right)-\frac{n_f^2}{108},
\eeq
i.e. $K_3=A_i^{(3)}/C_i$,
and $C^{(3)}$ has a long expression which can be found in Eq. (2.13) of Ref. \cite{NK3loopcusp}.

\subsection{Case with two massive eikonal lines}

We now give explicit expressions for the cusp anomalous dimension at one, two, and three loops, when both eikonal lines represent massive quarks that have the same mass $m$ (at the end of this subsection we will also consider two different masses). This is clearly applicable to the production of a heavy quark-antiquark pair. Here $v_i\cdot v_j=1+\beta^2$ and $v_i^2=v_j^2=1-\beta^2$, where $\beta=\sqrt{1-4m^2/s}$ is the speed of the massive quarks, and the cusp angle has the explicit form $\theta=\ln[(1+\beta)/(1-\beta)]$; also, we have $\beta=\tanh(\theta/2)$. 

We note that $\coth\theta=(1+\beta^2)/(2\beta)$, and we define 
\beq
L_{\beta}=\frac{(1+\beta^2)}{2\beta}\ln\left(\frac{1-\beta}{1+\beta}\right) \, .
\label{Lb}
\eeq
The cusp anomalous dimension at one loop is given by
\beq
\Gamma^{\beta\, (1)}_{\rm cusp}=-C_F \left( L_{\beta} +1 \right)
\label{cusp1loopb}
\eeq 
where we have added a superscript $\beta$ to indicate that the cusp anomalous dimension is here given in terms of $\beta$, with both eikonal lines having mass $m$. 

The cusp anomalous dimension at two loops is given by
\beq
\Gamma^{\beta \, (2)}_{\rm cusp}=K_2 \, \Gamma^{\beta\, (1)}_{\rm cusp}+C^{\beta \, (2)}
\label{cusp2loopb}
\eeq
where $C^{\beta \, (2)}=C_F C_A C^{' (2)}_{\beta}$,
with \cite{NK2loop}
\beqa
C^{' (2)}_{\beta}&=&\frac{1}{2}+\frac{\zeta_2}{2}
+\frac{1}{2}\ln^2\left(\frac{1-\beta}{1+\beta}\right) 
\nonumber \\ &&  \hspace{-3mm}
{}+\frac{(1+\beta^2)}{4\beta}\left[\zeta_2\ln\left(\frac{1-\beta}{1+\beta}\right)-\ln^2\left(\frac{1-\beta}{1+\beta}\right)
+\frac{1}{3}\ln^3\left(\frac{1-\beta}{1+\beta}\right)
-{\rm Li}_2\left(\frac{4\beta}{(1+\beta)^2}\right)\right] 
\nonumber \\ &&  \hspace{-3mm}
{}+\frac{(1+\beta^2)^2}{8\beta^2}\left[-\zeta_3-\zeta_2\ln\left(\frac{1-\beta}{1+\beta}\right)-\frac{1}{3}\ln^3\left(\frac{1-\beta}{1+\beta}\right)
-\ln\left(\frac{1-\beta}{1+\beta}\right) {\rm Li}_2\left(\frac{(1-\beta)^2}{(1+\beta)^2}\right) \right.
\nonumber \\ &&  \hspace{22mm} \left.
{}+{\rm Li}_3\left(\frac{(1-\beta)^2}{(1+\beta)^2}\right)\right]\, . 
\label{C2pb}
\eeqa

The cusp anomalous dimension at three loops is given by 
\beq
\Gamma^{\beta\, (3)}_{\rm cusp}= K_3 \, \Gamma^{\beta\, (1)}_{\rm cusp}
+2 K_2 \left(\Gamma^{\beta\, (2)}_{\rm cusp}-K_2 \, \Gamma^{\beta\, (1)}_{\rm cusp}\right)
+C^{\beta \, (3)} 
=K_3 \, \Gamma^{\beta\, (1)}_{\rm cusp}
+2 K_2 \, C^{\beta \, (2)}+C^{\beta \, (3)} 
\, ,
\eeq
where $C^{\beta\, (3)}=C_F C_A^2 C^{' (3)}_{\beta}$ with 
\beqa
C^{' (3)}_{\beta}&=&\frac{\zeta_2}{2}-\frac{\zeta_3}{8}
-\frac{9}{8}\zeta_4
+\frac{\zeta_2}{2}\ln\left(\frac{1-\beta}{1+\beta}\right)
-\frac{1}{4}\ln^2\left(\frac{1-\beta}{1+\beta}\right)
+\frac{1}{12}\ln^3\left(\frac{1-\beta}{1+\beta}\right)
-\frac{1}{24}\ln^4\left(\frac{1-\beta}{1+\beta}\right) 
\nonumber \\ && 
{}+\frac{1}{4}\ln^2\left(\frac{1-\beta}{1+\beta}\right)\ln \left(\frac{4\beta}{(1+\beta)^2}\right)
+\frac{3}{4} \ln\left(\frac{1-\beta}{1+\beta}\right) {\rm Li}_2\left(\frac{(1-\beta)^2}{(1+\beta)^2}\right)
-\frac{5}{8} {\rm Li}_3\left(\frac{(1-\beta)^2}{(1+\beta)^2}\right)
\nonumber \\ && \hspace{-13mm}
{}+\frac{(1+\beta^2)}{2\beta} \left\{-\frac{\zeta_3}{4}+\frac{15}{8}\zeta_4
-\left(\frac{\zeta_2}{2}-\frac{\zeta_3}{2}
+\frac{9}{8}\zeta_4\right) \ln\left(\frac{1-\beta}{1+\beta}\right)  
+\left(\frac{1}{4}+\zeta_2\right) \ln^2\left(\frac{1-\beta}{1+\beta}\right)
\right.
\nonumber \\ && \quad \quad 
{}-\left(\frac{1}{12}+\frac{\zeta_2}{3}\right)\ln^3\left(\frac{1-\beta}{1+\beta}\right)
+\frac{7}{24}\ln^4\left(\frac{1-\beta}{1+\beta}\right)
-\frac{1}{24}\ln^5\left(\frac{1-\beta}{1+\beta}\right)
\nonumber \\ && \quad \quad  
{}+\frac{1}{2}\ln^2\left(\frac{1-\beta}{1+\beta}\right) \ln \left(\frac{4\beta}{(1+\beta)^2}\right)
-\frac{1}{2}\ln^3\left(\frac{1-\beta}{1+\beta}\right) \ln \left(\frac{4\beta}{(1+\beta)^2}\right)
\nonumber \\ && \quad \quad  
{}-\frac{3}{4} \ln^2\left(\frac{1-\beta}{1+\beta}\right) {\rm Li}_2\left(\frac{(1-\beta)^2}{(1+\beta)^2}\right)
+\frac{1}{4} {\rm Li}_2\left(\frac{4\beta}{(1+\beta)^2}\right)
+\frac{1}{4} {\rm Li}_3\left(\frac{(1-\beta)^2}{(1+\beta)^2}\right)
\nonumber \\ && \quad \quad   \left.
{}+\frac{7}{4} \ln\left(\frac{1-\beta}{1+\beta}\right) {\rm Li}_3\left(\frac{(1-\beta)^2}{(1+\beta)^2}\right)
+\frac{1}{2} {\rm Li}_3\left(\frac{4\beta}{(1+\beta)^2}\right)-\frac{15}{8} {\rm Li}_4\left(\frac{(1-\beta)^2}{(1+\beta)^2}\right) \right\}
\nonumber \\ && \hspace{-13mm}
+\frac{(1+\beta^2)^2}{4\beta^2} \left\{-\frac{\zeta_2 \zeta_3}{2}-\frac{19}{8}\zeta_4+\frac{3}{2}\zeta_5
-\left(\frac{3}{2} \zeta_3-\frac{15}{8} \zeta_4 \right) \ln\left(\frac{1-\beta}{1+\beta}\right)
+\left(\frac{\zeta_3}{4}-\zeta_2\right) \ln^2\left(\frac{1-\beta}{1+\beta}\right)
\right.
\nonumber \\ && \quad \quad 
{}+\frac{2}{3} \zeta_2 \ln^3\left(\frac{1-\beta}{1+\beta}\right)
-\frac{1}{4}\ln^4\left(\frac{1-\beta}{1+\beta}\right)+\frac{11}{120}\ln^5\left(\frac{1-\beta}{1+\beta}\right)
\nonumber \\ && \quad \quad 
{}+\ln\left(\frac{4\beta}{(1+\beta)^2}\right)\left[\zeta_3+\zeta_2 \ln\left(\frac{1-\beta}{1+\beta}\right)
-\zeta_2 \ln^2\left(\frac{1-\beta}{1+\beta}\right) \right.
\nonumber \\ && \hspace{43mm} \left.
{}+\frac{1}{3}\ln^3\left(\frac{1-\beta}{1+\beta}\right)-\frac{1}{6}\ln^4\left(\frac{1-\beta}{1+\beta}\right)\right]
\nonumber \\ && \quad \quad 
{}-\ln^2\left(\frac{1-\beta}{1+\beta}\right) \ln^2\left(\frac{4\beta}{(1+\beta)^2}\right)
+\ln\left(\frac{1-\beta}{1+\beta}\right)  \ln^3\left(\frac{4\beta}{(1+\beta)^2}\right)
-\frac{1}{8} \ln^4\left(\frac{4\beta}{(1+\beta)^2}\right)
\nonumber \\ && \quad \quad
{}+\left[\frac{\zeta_2}{2}-\zeta_2\ln\left(\frac{1-\beta}{1+\beta}\right)
-2\ln^2\left(\frac{1-\beta}{1+\beta}\right)
-\frac{1}{12}\ln^3\left(\frac{1-\beta}{1+\beta}\right) \right.
\nonumber \\ && \hspace{18mm} \left.
{}+\ln\left(\frac{1-\beta}{1+\beta}\right) 
\ln\left(\frac{4\beta}{(1+\beta)^2}\right) \right]
{\rm Li}_2\left(\frac{(1-\beta)^2}{(1+\beta)^2}\right)
\nonumber \\ && \quad \quad
{}-\frac{1}{4} {\rm Li}_2^2\left(\frac{(1-\beta)^2}{(1+\beta)^2}\right)
+\frac{1}{2} \ln^2\left(\frac{4\beta}{(1+\beta)^2}\right) {\rm Li}_2\left(\frac{4\beta}{(1+\beta)^2}\right)
\nonumber \\ && \quad \quad
{}+\frac{1}{4} {\rm Li}_2^2\left(\frac{4\beta}{(1+\beta)^2}\right)
-\frac{1}{2} \ln^2\left(\frac{4\beta}{(1-\beta)^2}\right)
{\rm Li}_2\left(\frac{-(1-\beta)^2}{4\beta}\right) 
\nonumber \\ && \quad \quad
{}+\left[\frac{\zeta_2}{2}+\frac{3}{2}\ln\left(\frac{1-\beta}{1+\beta}\right)  
-\frac{1}{4}\ln^2\left(\frac{1-\beta}{1+\beta}\right)
-\ln\left(\frac{4\beta}{(1+\beta)^2}\right)\right]{\rm Li}_3\left(\frac{(1-\beta)^2}{(1+\beta)^2}\right)
\nonumber \\ && \quad \quad
{}+\left[\ln\left(\frac{1-\beta}{1+\beta}\right)
-\ln\left(\frac{4\beta}{(1+\beta)^2}\right)\right] 
{\rm Li}_3\left(\frac{4\beta}{(1+\beta)^2}\right)
\nonumber \\ && \quad \quad
{}+\left[2\ln\left(\frac{1-\beta}{1+\beta}\right)
-\ln \left(\frac{4\beta}{(1+\beta)^2}\right) \right] 
{\rm Li}_3\left(\frac{-(1-\beta)^2}{4\beta}\right)+\frac{9}{8} \ln\left(\frac{1-\beta}{1+\beta}\right)  {\rm Li}_4\left(\frac{(1-\beta)^2}{(1+\beta)^2}\right)
\nonumber \\ && \quad \quad \left.
{}+{\rm Li}_4\left(\frac{4\beta}{(1+\beta)^2}\right)-{\rm Li}_4\left(\frac{-(1-\beta)^2}{4\beta}\right)-\frac{3}{2} {\rm Li}_5\left(\frac{(1-\beta)^2}{(1+\beta)^2}\right) \right\}
\nonumber \\ && \hspace{-13mm} 
{}+\frac{1}{4} \left\{A(\beta)-A(0)+B(\beta)-B(0)\right\} 
\label{C3pb}
\eeqa
with 
\beqa
A(\beta)\!\!\!&=&\!\!\!
\frac{(1+\beta^2)^3}{8\beta^3} \left\{-3 \zeta_5
-4 \zeta_4 \ln\left(\frac{1-\beta}{1+\beta}\right)
-3 \zeta_3 \ln^2\left(\frac{1-\beta}{1+\beta}\right)
-\frac{4}{3} \zeta_2 \ln^3\left(\frac{1-\beta}{1+\beta}\right)
-\frac{1}{5} \ln^5\left(\frac{1-\beta}{1+\beta}\right)\right.
\nonumber \\ && \hspace{-10mm}
{}-\frac{2}{3} \ln^3\left(\frac{1-\beta}{1+\beta}\right) {\rm Li}_2\left(\frac{(1-\beta)^2}{(1+\beta)^2}\right) 
+\ln^2\left(\frac{1-\beta}{1+\beta}\right){\rm Li}_3\left(\frac{(1-\beta)^2}{(1+\beta)^2}\right)   
-2 \ln\left(\frac{1-\beta}{1+\beta}\right) {\rm Li}_4\left(\frac{(1-\beta)^2}{(1+\beta)^2}\right)
\nonumber \\ && \hspace{-10mm} \left.
{}+3 {\rm Li}_5\left(\frac{(1-\beta)^2}{(1+\beta)^2}\right) 
+H_{1,1,0,0,1}\left(\frac{4\beta}{(1+\beta)^2}\right)+H_{1,0,1,0,1}\left(\frac{4\beta}{(1+\beta)^2}\right) \right\} \, ,
\eeqa
where $A(0)=2$ and explicit expressions for the harmonic polylogarithms \cite{ERJV} $H_{1,1,0,0,1}$ and $H_{1,0,1,0,1}$ can be found in the Appendix of Ref. \cite{NK3loopcusp}, and 
\beqa
B(\beta)\!\!\!&=&\!\!\!
\frac{(1-\beta^2)}{4\beta}
\left\{-2 \zeta_2 \zeta_3 + 2 \zeta_3 \ln^2\left(\frac{1-\beta}{1+\beta}\right) 
+\left[\frac{3}{2} \zeta_4 -\frac{1}{6} \ln^4\left(\frac{1-\beta}{1+\beta}\right)\right]
\ln\left(\frac{2\beta}{1-\beta}\right) \right.
\nonumber \\ && \hspace{-16mm}
{}+\left[\frac{3}{2} \zeta_4-2\zeta_3 \ln\left(\frac{1-\beta}{1+\beta}\right) 
-\frac{1}{6} \ln^4\left(\frac{1-\beta}{1+\beta}\right)\right]
\ln\left(\frac{1-\beta}{2}\right)  
+2 \zeta_3 \left[{\rm Li}_2\left(\frac{-1+\beta}{1+\beta}\right)
+{\rm Li}_2\left(\frac{2\beta}{1+\beta}\right)\right]
\nonumber \\ && \hspace{-16mm} 
{}-\frac{2}{3} \ln^3\left(\frac{1-\beta}{1+\beta}\right)
\left[{\rm Li}_2\left(\frac{1-\beta}{1+\beta}\right)
-{\rm Li}_2\left(\frac{-1+\beta}{1+\beta}\right)\right]
+2 \ln^2\left(\frac{1-\beta}{1+\beta}\right) \left[{\rm Li}_3\left(\frac{1-\beta}{1+\beta}\right)
-{\rm Li}_3\left(\frac{-1+\beta}{1+\beta}\right)\right]
\nonumber \\ &&  \hspace{-16mm}
{}-4 \ln\left(\frac{1-\beta}{1+\beta}\right) 
\left[{\rm Li}_4\left(\frac{1-\beta}{1+\beta}\right)
-{\rm Li}_4\left(\frac{-1+\beta}{1+\beta}\right)\right]
+4 {\rm Li}_5\left(\frac{1-\beta}{1+\beta}\right)
-4 {\rm Li}_5\left(\frac{-1+\beta}{1+\beta}\right)
\nonumber \\ &&  \hspace{-16mm} \left. 
{}+4 \left[H_{1,0,1,0,0}\left(\frac{1-\beta}{1+\beta}\right)+H_{-1,0,1,0,0}\left(\frac{1-\beta}{1+\beta}\right)-H_{1,0,-1,0,0}\left(\frac{1-\beta}{1+\beta}\right)-H_{-1,0,-1,0,0}\left(\frac{1-\beta}{1+\beta}\right) \right] \right\}\!,
\label{cusp3loopb}
\eeqa
where $B(0)=3\zeta_3/2$ and explicit expressions for the harmonic polylogarithms, $H_{1,0,1,0,0}$, $H_{-1,0,1,0,0}$, $H_{1,0,-1,0,0}$, and $H_{-1,0,-1,0,0}$ can be found in the Appendix of Ref. \cite{NK3loopcusp}. 

The limit of the $n$-loop cusp anomalous dimension as $\beta$ goes to 1, or equivalently as $m^2/s\rightarrow 0$, can be written as 
\beq
\Gamma_{\rm cusp}^{\beta\, (n)}\stackrel{\beta\rightarrow 1}{\longrightarrow}  
-K_n C_F \ln\left(\frac{m^2}{s}\right) +R_n 
\label{beta1limit}
\eeq
where we define $K_n=A_i^{(n)}/C_i$ (and thus $K_1=1$), and the terms 
$R_n$ at one, two, and three loops are given, 
respectively, by 
$R_1=-K_1 C_F$, $R_2=-K_2 C_F+(1/2) C_F C_A (1-\zeta_3)$,
and 
\beq
R_3=-K_3 C_F+K_2 C_F C_A (1-\zeta_3)
+C_F C_A^2 \left(-\frac{1}{2}+\frac{3}{4}\zeta_2-\frac{\zeta_3}{4}
-\frac{3}{4}\zeta_2 \zeta_3+\frac{9}{8} \zeta_5 \right) \, .
\label{R3}
\eeq

Finally, we consider the case when the two lines have different masses, $m_1$ and $m_2$. In that case, we define
\beq
L_{\beta_1\beta_2}=\frac{(1+\beta_1\beta_2)}{2(\beta_1+\beta_2)}\ln\left(\frac{(1-\beta_1)(1-\beta_2)}{(1+\beta_1)(1+\beta_2)}\right) \, ,
\eeq
where $\beta_1=\left(1-\frac{4m_1^2 s}{(s+m_1^2-m_2^2)^2}\right)^{1/2}$ 
and
$\beta_2=\left(1-\frac{4m_2^2 s}{(s+m_2^2-m_1^2)^2}\right)^{1/2}$,  
and the cusp anomalous dimension at one loop is then given by
\beq
\Gamma^{\beta_1\beta_2\, (1)}_{\rm cusp}=-C_F \left(L_{\beta_1\beta_2} +1 \right) \, .
\eeq 

\subsection{Case with one massless and one massive eikonal line}

We take the eikonal line $i$ to represent a massive quark of mass $m_i$, and the eikonal line $j$ to represent a massless quark. To find the expressions for this case, we take the limit of the massive expression as the mass of eikonal line $j$ goes to zero, taking into account the self energies. We note that in this limit $\theta=\ln[2 p_i \cdot p_j/(m_i\sqrt{p_j^2})]$. At one loop, the heavy quark self-energy is -$C_F/2$, so removing it adds $C_F/2$, and then we add the massless contribution $C_F \ln\sqrt{p_j^2/s}$.  The overall change in the self-energy contributions relative to the fully massive case is thus $C_F/2+C_F \ln\sqrt{p_j^2/s}$, which equals $-R_1/2+C_F \ln\sqrt{p_j^2/s}$, where $R_1$ was defined in the previous subsection. Thus, we find at one loop 
\beq
\Gamma_{\rm cusp}^{m_i\, (1)}=C_F \left[\ln\left(\frac{2 p_i \cdot p_j}{m_i \sqrt{s}}\right) -\frac{1}{2}\right] \, ,
\label{1loop1}
\eeq  
where we have added a superscript $m_i$ to indicate that only eikonal line $i$ has a mass.

At two loops, the heavy-quark self-energy is $-(K_2/2) C_F +C_F C_A(1-\zeta_2)/4$, so again we remove it and then add the contribution for the 
massless eikonal line, which is $C_F K_2 \ln\sqrt{p_j^2/s}-C_F C_A(\zeta_2-\zeta_3)/4$. The total additional contribution is thus $(K_2/2) C_F -C_F C_A(1-\zeta_3)/4+C_F K_2 \ln\sqrt{p_j^2/s}$, which equals $-R_2/2+C_F K_2 \ln\sqrt{p_j^2/s}$, where $R_2$ was defined in the previous subsection.
Thus, we find the two-loop result
\beq
\Gamma_{\rm cusp}^{m_i \, (2)}=K_2 \, \Gamma_{\rm cusp}^{m_i\, (1)} +\frac{1}{4} C_F C_A (1-\zeta_3) \, .
\label{2loop1}
\eeq 
We note that this can be rewritten as 
$\Gamma_{\rm cusp}^{m_i \, (2)}=K_2 C_F\ln(2 p_i \cdot p_j/(m_i \sqrt{s}))+R_2/2$.

At three loops, again removing the heavy-quark self-energy and adding the massless contribution, we find a total additional contribution of 
$-R_3/2+C_F K_3 \ln\sqrt{p_j^2/s}$, where $R_3$ was defined in the previous subsection in Eq. (\ref{R3}).
Thus, we find the three-loop result 
\beq
\Gamma_{\rm cusp}^{m_i\, (3)}=K_3 \, \Gamma_{\rm cusp}^{m_i\, (1)} +\frac{1}{2}K_2 C_F C_A (1-\zeta_3)
+C_F C_A^2\left(-\frac{1}{4}+\frac{3}{8}\zeta_2-\frac{\zeta_3}{8}
-\frac{3}{8}\zeta_2 \zeta_3+\frac{9}{16} \zeta_5\right) \, .
\label{3loop1}
\eeq
We note that this can be rewritten as 
$\Gamma_{\rm cusp}^{m_i \, (3)}=K_3 C_F\ln(2 p_i \cdot p_j/(m_i \sqrt{s}))+R_3/2$.

We also note that in general the $n$-loop result can be written as 
$\Gamma_{\rm cusp}^{m_i \, (n)}=K_n C_F\ln(2 p_i \cdot p_j/(m_i \sqrt{s}))+R_n/2$.

\subsection{Case with two massless eikonal lines}

In the case where both eikonal lines are massless, $\theta=\ln(2 p_i \cdot p_j/\sqrt{p_i^2 \, p_j^2})$, and again removing the heavy-quark self-energies and adding the massless contributions, we find the lightlike cusp anomalous dimension as 
\beq
\Gamma_{\rm cusp}^{\rm massless}=C_F \, \ln\left(\frac{2 p_i \cdot p_j}{s}\right)  \sum_{n=1}^{\infty} \left(\frac{\alpha_s}{\pi}\right)^n K_n \, .
\eeq 

\mysection{$\Gamma_{\!\! S}$ for some processes with trivial color structure}

The soft anomalous dimension vanishes to all orders for many processes with trivial color structure, with no colored particles in the final state. Some well-known processes of this type are the Drell-Yan processes $q{\bar q} \rightarrow \gamma^*$, $q{\bar q} \rightarrow Z$; $W$-boson production via $q{\bar q'} \rightarrow W^{\pm}$; Higgs production via $b{\bar b} \rightarrow H$ and $gg \rightarrow H$; production of electroweak boson pairs $q{\bar q} \rightarrow \gamma \gamma$, $q{\bar q} \rightarrow Z Z$,  $q{\bar q} \rightarrow W^+ W^-$, $q{\bar q} \rightarrow \gamma Z$; $q{\bar q'} \rightarrow W^{\pm} \gamma $; $q{\bar q'} \rightarrow W^{\pm} Z$; and charged Higgs production via $b{\bar b} \rightarrow H^- W^+$, $b{\bar b} \rightarrow H^+ H^-$, $gg \rightarrow H^+ H^-$.

The soft anomalous dimension for deep-inelastic-scattering (DIS) $lq\rightarrow lq$ with underlying process $q\gamma^* \rightarrow q$ is given at one, two, and three loops by 
\beq
\Gamma_{\!\! S \, q \gamma^* \rightarrow q}^{(1)}=C_F \ln(-t/s)\, , \quad 
\Gamma_{\!\! S\, q \gamma^* \rightarrow q}^{(2)}=K_2 C_F \ln(-t/s) \, , \quad 
\Gamma_{\!\! S\, q \gamma^* \rightarrow q}^{(3)}=K_3 C_F \ln(-t/s) \, ,
\eeq 
respectively.

\mysection{$\Gamma_{\!\! S}$ for large-$p_T$ $W$ production and related processes}

We next consider processes with a $W$-boson or a $Z$-boson or a Higgs-boson produced at large-$p_T$, direct-photon production, and related processes. 
For these processes there is only a single color tensor, coupling the two quarks (or quark and antiquark) to the gluon in an octet state. Thus the soft anomalous dimension is a simple function, not a matrix.

The soft anomalous dimensions for the processes $qg\rightarrow W^{\pm}q'$ and $qg\rightarrow Zq$ and $qg \rightarrow \gamma q$ and $bg \rightarrow Hb$, are all identical. Using $W$ production at large $p_T$ as the specific process, the soft anomalous dimension is given at one loop by \cite{LOS,NKun04,NKRG}
\beq
\Gamma_{\!\! S \, qg\rightarrow Wq'}^{(1)}=C_F \ln\left(\frac{-u}{s}\right)
+\frac{C_A}{2} \ln\left(\frac{t}{u}\right) \, ,
\eeq
at two-loops by \cite{NKRG}
\beq
\Gamma_{\!\! S \, qg \rightarrow Wq'}^{(2)}=K_2 \, \Gamma_{\!\! S \, qg \rightarrow Wq'}^{(1)} \, ,
\eeq
and at three loops by
\beq
\Gamma_{\!\! S \, qg \rightarrow Wq'}^{(3)}=K_3 \, \Gamma_{\!\! S \, qg \rightarrow Wq'}^{(1)} \, .
\eeq

For $q {\bar q'}\rightarrow W^{\pm}g$ or  $q {\bar q}\rightarrow Zg$ or  $q{\bar q} \rightarrow  \gamma g$ or $b{\bar b} \rightarrow Hg$, the corresponding results are
\beq
\Gamma_{\!\! S \, q{\bar q'}\rightarrow Wg}^{(1)}=\frac{C_A}{2} \ln\left(\frac{tu}{s^2}\right)\, , \quad
\Gamma_{\!\! S \, q{\bar q'} \rightarrow Wg}^{(2)}=K_2 \, \Gamma_{\!\! S \, q{\bar q'} \rightarrow Wg}^{(1)} \, , \quad
\Gamma_{\!\! S \, q{\bar q'} \rightarrow Wg}^{(3)}=K_3 \, \Gamma_{\!\! S \, q{\bar q'} \rightarrow Wg}^{(1)} \, .
\eeq

We also note that the soft anomalous dimensions for the reverse processes $\gamma q \rightarrow qg$ and $\gamma g \rightarrow q {\bar q}$ are the same as for the corresponding processes above.

\mysection{$\Gamma_{\!\! S}$ for single-top production and related processes}

We continue with various single-top production processes. They include $s$-channel, $t$-channel, and $tW^-$ production, and various related FCNC single-top processes.

\subsection{$s$-channel single-top production}

Soft anomalous dimensions for $s$-channel single-top production were calculated at one loop in Refs. \cite{NKst,NKsch,NKtoprev}, at two loops in \cite{NKsch,NKtop3loop}, and at three loops in \cite{NKtop3loop}.

The partonic processes are $q{\bar q}' \rightarrow t{\bar b}$.
In this channel we have $2 \rightarrow 2$ processes at lowest order that involve a final-state top quark and a final-state massless quark. Thus, we have four colored particles involved in the scattering, one of which is massive.

The color structure of the hard scattering in $s$-channel single-top production is more complicated and thus the soft anomalous dimension is a $2 \times 2$ matrix in color space. 
We choose a singlet-octet $s$-channel color basis, $c_1^{q{\bar q}' \rightarrow t {\bar b}}=\delta_{ab} \delta_{12}$ and $c_2^{q{\bar q}' \rightarrow t {\bar b}}=T^c_{ba} T^c_{12}$. 

The four matrix elements of the $s$-channel soft anomalous dimension matrix, $\Gamma_{\!\! S \, q{\bar q}' \rightarrow t {\bar b}}$, are given at one loop by 
\beqa
\Gamma_{\!\! 11 \, q{\bar q}' \rightarrow t {\bar b}}^{(1)}\!\!\!\!&=&\!\!\!C_F \left[\ln\left(\frac{s-m_t^2}{m_t\sqrt{s}}\right)-\frac{1}{2}\right], \,  
\Gamma_{\!\! 12 \, q{\bar q}' \rightarrow t {\bar b}}^{(1)}\!\!=\frac{C_F}{2N_c} \ln\left(\frac{t(t-m_t^2)}{u(u-m_t^2)}\right), \,
\Gamma_{\!\! 21 \, q{\bar q}' \rightarrow t {\bar b}}^{(1)}\!\!=\ln\left(\frac{t(t-m_t^2)}{u(u-m_t^2)}\right),
\nonumber \\
\Gamma_{\!\! 22 \, q{\bar q}' \rightarrow t {\bar b}}^{(1)}\!\!\!&=&\!\!\!\left(C_F-\frac{C_A}{2}\right) \left[\ln\left(\frac{s-m_t^2}{m_t \sqrt{s}}\right)-\frac{1}{2}+2\ln\left(\frac{t(t-m_t^2)}{u(u-m_t^2)}\right)\right]
+\frac{C_A}{2} \left[\ln\left(\frac{t(t-m_t^2)}{m_t \, s^{3/2}}\right)-\frac{1}{2}\right]
\label{Gamma1s}
\eeqa
where $m_t$ is the top-quark mass.

At two loops, we have
\beqa
\Gamma_{\!\! 11 \, q{\bar q}' \rightarrow t {\bar b}}^{(2)}&=& K_2 \, \Gamma_{\!\! 11 \, q{\bar q}' \rightarrow t {\bar b}}^{(1)}+\frac{1}{4} C_F C_A (1-\zeta_3) \, , \quad \quad
\Gamma_{\!\! 12 \, q{\bar q}' \rightarrow t {\bar b}}^{(2)}= K_2 \, \Gamma_{\!\! 12 \, q{\bar q}' \rightarrow t {\bar b}}^{(1)} \, ,
\nonumber \\
\Gamma_{\!\! 21 \, q{\bar q}' \rightarrow t {\bar b}}^{(2)}&=& K_2 \, \Gamma_{\!\! 21 \, q{\bar q}' \rightarrow t {\bar b}}^{(1)}, \quad \quad
\Gamma_{\!\! 22 \, q{\bar q}' \rightarrow t {\bar b}}^{(2)}= K_2 \, \Gamma_{\!\! 22 \, q{\bar q}' \rightarrow t {\bar b}}^{(1)}+\frac{1}{4} C_F C_A (1-\zeta_3) \, .
\label{Gamma2s}
\eeqa

At three loops, only the first element of the soft anomalous dimension matrix is needed to calculate the N$^3$LO soft-gluon corrections, because only the first element of the leading-order hard matrix is nonzero. We have
\beq
\Gamma_{\!\! 11 \, q{\bar q}' \rightarrow t {\bar b}}^{(3)}= K_3 \, \Gamma_{\!\! 11 \, q{\bar q}' \rightarrow t {\bar b}}^{(1)}
+\frac{1}{2} K_2 C_F C_A (1-\zeta_3)
+C_F C_A^2\left(-\frac{1}{4}+\frac{3}{8}\zeta_2-\frac{\zeta_3}{8}-\frac{3}{8}\zeta_2 \zeta_3+\frac{9}{16} \zeta_5\right) \, .
\label{Gs113}
\eeq
Furthermore, up to unknown contributions from four-parton correlations, $\Gamma_{\!\! 22 \, q{\bar q}' \rightarrow t {\bar b}}^{(3)}$ should have the same form as Eq. (\ref{Gs113}) (just replace the 11 subscripts by 22), and the two off-diagonal elements at three loops should have a similar form to Eq. (\ref{Gamma2s}) (replace $K_2$ by $K_3$). 

\subsection{$t$-channel single-top production}

Soft anomalous dimensions for $t$-channel single-top production were calculated at one loop in Refs. \cite{NKst,NKtch,NKtoprev}, at two loops in \cite{NKtch,NKtop3loop}, and at three loops in \cite{NKtop3loop}.

The partonic processes are $bq \rightarrow tq'$.
The color structure of the hard scattering in $t$-channel single-top production is again complicated, and 
the soft anomalous dimension is a $2 \times 2$ matrix in color space. 
We choose a singlet-octet $t$-channel color basis, $c_1^{bq \rightarrow tq'}=\delta_{a1} \delta_{b2}$ 
and $c_2^{bq \rightarrow tq'}=T^c_{1a} T^c_{2b}$.

The four matrix elements of the $t$-channel soft anomalous dimension matrix, ${\Gamma}_{\!\! S \, bq \rightarrow tq'}$, are given at one loop by 
\beqa
{\Gamma}_{\!\! 11 \, bq \rightarrow tq'}^{(1)}\!\!\!&=&\!\!\!
C_F \left[\ln\left(\frac{t(t-m_t^2)}{m_t s^{3/2}}\right)-\frac{1}{2}\right], \,
{\Gamma}_{\!\! 12 \, bq \rightarrow tq'}^{(1)}\!\!=\frac{C_F}{2N_c} \ln\left(\frac{u(u-m_t^2)}{s(s-m_t^2)}\right),
{\Gamma}_{\!\! 21 \, bq \rightarrow tq'}^{(1)}\!\!=\ln\left(\frac{u(u-m_t^2)}{s(s-m_t^2)}\right) \, ,
\nonumber \\
{\Gamma}_{\!\! 22 \, bq \rightarrow tq'}^{(1)}\!\!\!&=&\!\!\! \left(C_F-\frac{C_A}{2}\right) \left[\ln\left(\frac{t(t-m_t^2)}{m_t \, s^{3/2}}\right)-\frac{1}{2}
+2\ln\left(\frac{u(u-m_t^2)}{s(s-m_t^2)}\right)\right] 
+\frac{C_A}{2}\ln\left[\left(\frac{u(u-m_t^2)}{m_t \, s^{3/2}}\right)-\frac{1}{2}\right].
\nonumber \\
\label{Gamma1t}
\eeqa

At two loops,  our calculation gives 
\beqa
\Gamma_{\!\! 11 \, bq \rightarrow tq'}^{(2)}&=& K_2 \, \Gamma_{\!\! 11 \, bq \rightarrow tq'}^{(1)}+\frac{1}{4} C_F C_A (1-\zeta_3) \, , \quad \quad
\Gamma_{\!\! 12 \, bq \rightarrow tq'}^{(2)}= K_2 \, \Gamma_{\!\! 12 \, bq \rightarrow tq'}^{(1)} \, ,
\nonumber \\
\Gamma_{\!\! 21 \, bq \rightarrow tq'}^{(2)}&=& K_2 \, \Gamma_{\!\! 21 \, bq \rightarrow tq'}^{(1)} \, ,
\quad \quad
\Gamma_{\!\! 22 \, bq \rightarrow tq'}^{(2)}= K_2 \, \Gamma_{\!\! 22 \, bq \rightarrow tq'}^{(1)}+\frac{1}{4} C_F C_A (1-\zeta_3) \, .
\label{Gamma2t}
\eeqa

At three loops, only the first element of the soft anomalous dimension matrix is needed to calculate the N$^3$LO soft-gluon corrections, as for the $s$-channel. We find
\beq
\Gamma_{\!\! 11 \, bq \rightarrow tq'}^{(3)}= K_3 \, \Gamma_{\!\! 11 \, bq \rightarrow tq'}^{(1)}
+\frac{1}{2} K_2 C_F C_A (1-\zeta_3)
+C_F C_A^2\left(-\frac{1}{4}+\frac{3}{8}\zeta_2-\frac{\zeta_3}{8}-\frac{3}{8}\zeta_2 \zeta_3+\frac{9}{16} \zeta_5\right) \, .
\label{Gt113}
\eeq
Furthermore, up to unknown contributions from four-parton correlations, $\Gamma_{\!\! 22 \, bq \rightarrow tq'}^{(3)}$ should have the same form as Eq. (\ref{Gt113}) (just replace the 11 subscripts by 22), and the two off-diagonal elements at three loops should have similar form to Eq. (\ref{Gamma2t}) (replace $K_2$ by $K_3$). 

\subsection{$bg \rightarrow tW^-$ and related processes}

We next present the soft anomalous dimension for the associated production of a top quark with a $W$ boson via $bg\rightarrow tW^-$ which is known at one-loop \cite{NKun04,NKst}, two loops \cite{NKtWH}, and three loops \cite{NKtop3loop}.  The soft anomalous dimension for $tW$ production is identical to that for other related processes in models of new physics, such as the associated production of a top quark with a charged Higgs boson via $bg\rightarrow tH^-$ in two-Higgs-doublet models, and flavor-changing-neutral-current (FCNC) processes that proceed via anomalous top-quark couplings, such $qg\rightarrow tZ$ (or $tZ')$ and $qg\rightarrow t\gamma$ (see \cite{NKtoprev} for a review). In all these cases we have $2 \rightarrow 2$ processes at lowest order that involve a final-state top quark and a final-state boson.

The soft anomalous dimension at one-loop for $bg \rightarrow tW^-$ (and all other processes in this set) is given by 
\beq
\Gamma_{\!\! S \, bg \rightarrow tW}^{(1)}=C_F \left[\ln\left(\frac{m_t^2-t}{m_t\sqrt{s}}\right)
-\frac{1}{2}\right] +\frac{C_A}{2} \ln\left(\frac{u-m_t^2}{t-m_t^2}\right) \, .
\eeq

The two-loop result is given by
\beq
\Gamma_{\!\! S \, bg \rightarrow tW}^{(2)}=K_2 \, \Gamma_{\!\! S \, bg \rightarrow tW}^{(1)}
+\frac{1}{4}C_F C_A (1-\zeta_3) \, .
\eeq

The three-loop result is
\beq
\Gamma_{\!\! S \, bg \rightarrow tW}^{(3)}=K_3 \, \Gamma_{\!\! S \, bg \rightarrow tW}^{(1)}+\frac{1}{2} K_2 C_F C_A (1-\zeta_3)+C_F C_A^2\left(-\frac{1}{4}+\frac{3}{8}\zeta_2-\frac{\zeta_3}{8}-\frac{3}{8}\zeta_2 \zeta_3+\frac{9}{16} \zeta_5\right) \, .
\eeq

\subsection{FCNC $ue \rightarrow te$}

For the FCNC process $ue \rightarrow te$,  
which proceeds via anomalous $t$-$q$-$\gamma$ and $t$-$q$-$Z$ couplings, 
we have \cite{ABNK,NKun04,NKAB}
\beq
\Gamma_{\!\!S \, ue \rightarrow te}^{(1)}=C_F \left[\ln\left(\frac{m_t^2-t}{m_t{\sqrt s}}\right)-\frac{1}{2}\right] \, .
\eeq
The two-loop result is given by
\beq
\Gamma_{\!\!S \, ue \rightarrow te}^{(2)}=K_2 \, \Gamma_{\!\!S \, ue \rightarrow te}^{(1)} +\frac{1}{4}C_F C_A (1-\zeta_3)
\eeq
and the three-loop result by
\beq
\Gamma_{\!\!S \, ue \rightarrow te}^{(3)}=K_3 \, \Gamma_{\!\!S \, ue \rightarrow te}^{(1)} +\frac{1}{2} K_2 C_F C_A (1-\zeta_3)
+C_F C_A^2\left(-\frac{1}{4}+\frac{3}{8}\zeta_2-\frac{\zeta_3}{8}-\frac{3}{8}\zeta_2 \zeta_3+\frac{9}{16} \zeta_5\right) \, .
\eeq

\subsection{FCNC $tg$ production}

We next consider $tg$ production via an anomalous $t$-$q$-$g$ coupling \cite{NKEM}.
For the partonic process $gu \rightarrow tg$ we choose the color basis 
$c_1^{gu \rightarrow tg} = \delta_{b1} \delta_{a2}$, $c_2^{gu \rightarrow tg} = d^{a2c} T^c_{1b}$, 
$c_3^{gu \rightarrow tg} = i f^{a2c} T^c_{1b}$.
Then the one-loop soft anomalous dimension is
\beq
\Gamma_{\!\! S \, gu \rightarrow tg}^{(1)} =  \left[\begin{array}{ccc}
\Gamma_{\!\! 11 \, gu \rightarrow tg}^{(1)} & 0 & \Gamma_{\!\! 13 \, gu \rightarrow tg}^{(1)} \\
0 & \Gamma_{\!\! 22 \, gu \rightarrow tg}^{(1)} & \Gamma_{\!\! 23 \, gu \rightarrow tg}^{(1)} \\
\Gamma_{\!\! 31 \, gu \rightarrow tg}^{(1)} & \Gamma_{\!\! 32 \, gu \rightarrow tg}^{(1)}  & \Gamma_{\!\! 22 \, gu \rightarrow tg}^{(1)}
\end{array}
\right] 
\eeq
where \cite{NKEM}
\beqa
\Gamma_{\!\! 11 \, gu \rightarrow tg}^{(1)} &=& C_F \left[\ln\left(\frac{m_t^2-u}{m_t\sqrt{s}} \right) - \frac{1}{2} \right] + C_A \ln\left(\frac{-u}{s} \right) \, , 
\nonumber \\
\Gamma_{\!\! 31 \, gu \rightarrow tg}^{(1)} &=& \ln \left(\frac{t(t-m_t^2)}{s(s-m_t^2)}\right) \, , \quad 
\Gamma_{\!\! 32 \, gu \rightarrow tg}^{(1)} = \frac{(N_c^2-4)}{4N_c} \ln \left(\frac{t(t-m_t^2)}{s(s-m_t^2)}\right) \, ,
\nonumber \\
\Gamma_{\!\! 22 \, gu \rightarrow tg}^{(1)} &=& C_F \left[\ln\left(\frac{m_t^2-u}{m_t\sqrt{s}}\right) - \frac{1}{2} \right] +  \frac{C_A}{4} \ln\left(\frac{t u^2 (s-m_t^2)(t-m_t^2)}{(u-m_t^2)^2 \, s^3}\right) \, ,
\nonumber \\
\Gamma_{\!\! 23 \, gu \rightarrow tg}^{(1)} &=& \frac{C_A}{4} \ln \left(\frac{t(t-m_t^2)}{s(s-m_t^2)}\right) \, , \quad
\Gamma_{\!\! 13 \, gu \rightarrow tg}^{(1)}=\frac{1}{2} \ln \left(\frac{t(t-m_t^2)}{s(s-m_t^2)}\right) \, .
\eeqa

For the two-loop soft anomalous dimension matrix, as for $s$-channel and $t$-channel single-top production in the previous subsections, the two-loop matrix elements are given by $K_2$ times the corresponding one-loop elements, with an additional term $C_FC_A(1-\zeta_3)/4$ in the diagonal elements.

\mysection{$\Gamma_{\!\! S}$ for $t{\bar t}$ production and related processes}

In this section we discuss the soft anomalous dimension matrices for top-antitop pair production, which of course are the same for bottom quark or charm quark production, and related processes such as DIS heavy-quark production, FCNC $tt$ production, and squark and gluino production. 

\subsection{$t{\bar t}$ production in hadronic collisions}

The top-antitop pair production partonic processes at lowest order are
$q{\bar q} \rightarrow t{\bar t}$ and $gg \rightarrow t{\bar t}$.
Next, we present the one-loop  and two-loop results for the soft anomalous matrices for these partonic processes \cite{NKGS1,NKGS2,NK2loop,FNPY1,FNPY2,NKnnll} as well as a form for the three-loop results. 

\subsubsection{$q{\bar q } \rightarrow t{\bar t}$}

The soft anomalous dimension matrix $\Gamma_{\!\! S \, q{\bar q}\rightarrow t{\bar t}}$ for the process $q{\bar q} \rightarrow t{\bar t}$ is a $2 \times 2$ matrix. We use a color tensor basis of $s$-channel singlet and octet exchange, 
$c_1^{ q{\bar q}\rightarrow t{\bar t}} = \delta_{ab}\delta_{12}$, 
$c_2^{ q{\bar q}\rightarrow t{\bar t}} =  T^c_{ba} \, T^c_{12}$.

The four matrix elements of $\Gamma_{\!\! S \, q{\bar q}\rightarrow t{\bar t}}$ are given at one loop \cite{NKGS1,NKGS2,NKnnll} by 
\beqa
\Gamma_{\!\! 11 \, q{\bar q}\rightarrow t{\bar t}}^{(1)}&=&=-C_F\left(L_{\beta}+1\right) \, ,
\quad
\Gamma_{\!\! 12 \, q{\bar q}\rightarrow t{\bar t}}^{(1)}=
\frac{C_F}{N_c} \ln\left(\frac{t-m_t^2}{u-m_t^2}\right) \, ,
\quad
\Gamma_{\!\! 21 \, q{\bar q}\rightarrow t{\bar t}}^{(1)}=
2\ln\left(\frac{t-m_t^2}{u-m_t^2}\right) \, ,
\nonumber \\ 
\Gamma_{\!\! 22 \, q{\bar q}\rightarrow t{\bar t}}^{(1)}&=&\left(C_F-\frac{C_A}{2}\right)
\left[-L_{\beta}-1+4\ln\left(\frac{t-m_t^2}{u-m_t^2}\right)\right]
+\frac{C_A}{2}\left[\ln\left(\frac{(t-m_t^2)^2}{s\, m_t^2}\right)-1\right] \, ,
\eeqa
where $L_{\beta}$ is given by Eq. (\ref{Lb}). We note that the first element of the matrix is equal to $\Gamma_{\rm cusp}^{\beta \, (1)}$, Eq. (\ref{cusp1loopb}).

At two loops we have \cite{NK2loop,FNPY1,FNPY2,NKnnll}
\beqa
\Gamma_{\!\! 11 \, q{\bar q}\rightarrow t{\bar t}}^{(2)}&=&\Gamma_{\rm cusp}^{\beta \, (2)} \, ,
\quad \Gamma_{\!\! 12 \, q{\bar q}\rightarrow t{\bar t}}^{(2)}=
\left(K_2-C_A N_2^{\beta}\right) \Gamma_{\!\! 12 \, q{\bar q}\rightarrow t{\bar t}}^{(1)} \, ,
\quad 
\Gamma_{\!\! 21 \, q{\bar q}\rightarrow t{\bar t}}^{(2)}=
\left(K_2+C_A N_2^{\beta}\right) \Gamma_{\!\! 21 \, q{\bar q}\rightarrow t{\bar t}}^{(1)} \, ,
\nonumber \\
\Gamma_{\!\! 22 \, q{\bar q}\rightarrow t{\bar t}}^{(2)}&=&
K_2 \, \Gamma_{\!\! 22 \, q{\bar q}\rightarrow t{\bar t}}^{(1)}
+C_A\left(C_F-\frac{C_A}{2}\right) C^{' (2)}_{\beta} +\frac{C_A^2}{4}(1-\zeta_3)\, ,
\label{Gqqtt2}
\eeqa
where $\Gamma_{\rm cusp}^{\beta \, (2)}$ is given by Eq. (\ref{cusp2loopb}), $C^{' (2)}_{\beta}$ is given by Eq. (\ref{C2pb}), and $N_2^{\beta}$ is given by 
\beqa
N_2^{\beta}&=&\frac{1}{4}\ln^2\left(\frac{1-\beta}{1+\beta}\right)
+\frac{(1+\beta^2)}{8 \beta} \left[\zeta_2
-\ln^2\left(\frac{1-\beta}{1+\beta}\right)
-{\rm Li}_2\left(\frac{4\beta}{(1+\beta)^2}\right)\right] \, .
\eeqa

At three loops, we expect a similar structure up to four-parton correlations, i.e. the first three matrix elements should have the same form as in Eq. (\ref{Gqqtt2}) (replace two-loop quantities by three-loop ones), while
\beqa
\Gamma_{\!\! 22 \, q{\bar q}\rightarrow t{\bar t}}^{(3)}&=&
K_3 \, \Gamma_{\!\! 22 \, q{\bar q}\rightarrow t{\bar t}}^{(1)}
+C_A\left(C_F-\frac{C_A}{2}\right) \left(C_A  C^{' (3)}_{\beta}+2 K_2 C^{' (2)}_{\beta}\right)+\frac{K_2}{2}C_A^2(1-\zeta_3)
\nonumber \\ &&
{}+C_A^3\left(-\frac{1}{4}+\frac{3}{8}\zeta_2-\frac{\zeta_3}{8}-\frac{3}{8}\zeta_2\zeta_3+\frac{9}{16}\zeta_5\right)+X_{\!\! 22 \, q{\bar q}\rightarrow t{\bar t}}^{(3)\,4p}\, ,
\label{Gqqtt223}
\eeqa
where $C^{' (3)}_{\beta}$ is given by Eq. (\ref{C3pb}), and $X_{\!\! 22 \, q{\bar q}\rightarrow t{\bar t}}^{(3)\,4p}$ denotes the unknown three-loop contributions from four-parton correlations.

\subsubsection{$gg \rightarrow t{\bar t}$}

The soft anomalous dimension matrix $\Gamma_{\!\! S \, gg\rightarrow t{\bar t}}$ for the process $gg \rightarrow t{\bar t}$ in a color tensor basis
$c_1^{gg\rightarrow t{\bar t}}=\delta^{ab}\,\delta_{12}$, $c_2^{gg\rightarrow t{\bar t}}=d^{abc}\,T^c_{12}$, $c_3^{gg\rightarrow t{\bar t}}=i f^{abc}\,T^c_{12}$, is given by 
\beqa
\Gamma_{\!\! S \, gg\rightarrow t{\bar t}}=\left[\begin{array}{ccc}
\Gamma_{\!\! 11 \, gg\rightarrow t{\bar t}} & 0 & \Gamma_{\!\! 13 \, gg\rightarrow t{\bar t}} \vspace{2mm} \\
0 & \Gamma_{\!\! 22 \, gg\rightarrow t{\bar t}} & \Gamma_{\!\! 23 \, gg\rightarrow t{\bar t}} \vspace{2mm} \\
\Gamma_{\!\! 31 \, gg\rightarrow t{\bar t}} & \Gamma_{\!\! 32 \, gg\rightarrow t{\bar t}} & \Gamma_{\!\! 22 \, gg\rightarrow t{\bar t}}
\end{array}
\right] \, .
\label{Gsggtt}
\eeqa

At one loop we have \cite{NKGS2,NKnnll}
\beqa
\Gamma_{\!\! 11 \, gg\rightarrow t{\bar t}}^{(1)}&=& -C_F\left(L_{\beta}+1\right)  \, ,
\quad
\Gamma_{\!\! 13 \, gg\rightarrow t{\bar t}}^{(1)}= \ln\left(\frac{t-m_t^2}{u-m_t^2}\right) \, , \quad 
\Gamma_{\!\! 31 \, gg\rightarrow t{\bar t}}^{(1)}= 2 \ln\left(\frac{t-m_t^2}{u-m_t^2}\right) \, , 
\nonumber \\
\Gamma_{\!\! 22 \, gg\rightarrow t{\bar t}}^{(1)}&=& \left(C_F-\frac{C_A}{2}\right) \left(-L_{\beta}-1\right)
+\frac{C_A}{2}\left[\ln\left(\frac{(t-m_t^2)(u-m_t^2)}{s\, m_t^2}\right)-1\right] \, ,
\nonumber \\
\Gamma_{\!\! 23 \, gg\rightarrow t{\bar t}}^{(1)}&=&\frac{C_A}{2} \ln\left(\frac{t-m_t^2}{u-m_t^2}\right) \, , \quad
\Gamma_{\!\! 32 \, gg\rightarrow t{\bar t}}^{(1)}=\frac{(N_c^2-4)}{2N_c} \ln\left(\frac{t-m_t^2}{u-m_t^2}\right) \, .
\eeqa

At two loops we find \cite{NK2loop,FNPY2,NKnnll}
\beqa
\Gamma_{\!\! 11 \, gg\rightarrow t{\bar t}}^{(2)}&=& \Gamma_{\rm cusp}^{\beta \, (2)} \, ,
\quad
\Gamma_{\!\! 13 \, gg\rightarrow t{\bar t}}^{(2)}=\left(K_2-C_A N_2^{\beta}\right) 
\Gamma_{\!\! 13 \, gg\rightarrow t{\bar t}}^{(1)} \, ,
\quad 
\Gamma_{\!\! 31 \, gg\rightarrow t{\bar t}}^{(2)}=\left(K_2+C_A N_2^{\beta}\right)  
\Gamma_{\!\! 31 \, gg\rightarrow t{\bar t}}^{(1)} \, ,
\nonumber \\
\Gamma_{\!\! 22 \, gg\rightarrow t{\bar t}}^{(2)}&=& K_2 \, \Gamma_{\!\! 22 \, gg\rightarrow t{\bar t}}^{(1)}
+C_A\left(C_F-\frac{C_A}{2}\right) C^{' (2)}_{\beta}+\frac{C_A^2}{4}(1-\zeta_3) \, ,
\nonumber \\
\Gamma_{\!\! 23 \, gg\rightarrow t{\bar t}}^{(2)}&=& K_2 \, \Gamma_{\!\! 23 \, gg\rightarrow t{\bar t}}^{(1)} \, , \quad
\Gamma_{\!\! 32 \, gg\rightarrow t{\bar t}}^{(2)}= K_2 \, \Gamma_{\!\! 32 \, gg\rightarrow t{\bar t}}^{(1)} \, . 
\eeqa

At three loops, we expect a similar structure but with additional four-parton correlations, as discussed in the previous subsection, i.e. replace two-loop quantities by three-loop ones, and $\Gamma_{\!\! 22 \, gg\rightarrow t{\bar t}}^{(3)}$ of the same general form as Eq. (\ref{Gqqtt223}).

\subsection{DIS heavy-quark production}

For heavy-quark production in DIS (also known as electroproduction of heavy quarks), $ep \rightarrow eQ{\bar Q}$, the underlying process is 
$g \gamma^* \rightarrow Q {\bar Q}$.
As for direct photon production there is only
a single color tensor, coupling the produced
pair to the gluon in an octet state. The soft anomalous dimension is given 
at one loop \cite{NKun04,ELSM} by
\beq
\Gamma_{\!\! S \, g \gamma^* \rightarrow Q{\bar Q}}^{(1)}=
-C_F(L_\beta + 1)+\frac{C_A}{2} \left[L_{\beta}+\ln\left(\frac{(t-m_Q^2)(u-m_Q^2)}{m_Q^2 \, s}\right) \right]  \, ,
\eeq
where $L_{\beta}$ is given by Eq. (\ref{Lb}).

\subsection{$e^+ e^- \rightarrow t{\bar t}$}

The soft anomalous dimension for $e^+ e^- \rightarrow t{\bar t}$ is simply the cusp anomalous dimension for the case of both eikonal lines of mass $m_t$ \cite{NK2loop,NK3loopcusp} that was presented in Section 4.1.

\subsection{FCNC $qq \rightarrow tt$}

For the process, $uu \rightarrow tt$ \cite{NKAB},
which proceeds via anomalous $t$-$q$-$\gamma$ and $t$-$q$-$Z$ couplings, 
we choose a color basis consisting of
singlet exchange in the $t$ and $u$ channels,
$c_1=\delta_{a1} \delta_{b2}$ and $c_2=\delta_{a2} \delta_{b1}$.
Then $\Gamma_{\!\! S \, qq \rightarrow tt}$ is a $2 \times 2$ soft anomalous dimension matrix, and its matrix elements at one loop \cite{NKAB} are given by 
\beqa
\Gamma_{\!\! 11 \, qq \rightarrow tt}^{(1)}&=&C_F\left[2\ln\left(\frac{m_t^2-t}{m_t{\sqrt s}}\right)-1\right]
+\left(C_F-\frac{C_A}{2}\right) \left[2\ln\left(\frac{m_t^2-u}{m_t{\sqrt s}}\right)
+L_{\beta}\right] \, ,
\nonumber \\
\Gamma_{\!\! 12 \, qq \rightarrow tt}^{(1)}&=&\ln\left(\frac{m_t^2-t}{m_t{\sqrt s}}\right)+\frac{1}{2}L_{\beta} \, , \quad \quad 
\Gamma_{\!\! 21 \, qq \rightarrow tt}^{(1)}=\ln\left(\frac{m_t^2-u}{m_t{\sqrt s}}\right)
+\frac{1}{2}L_{\beta} \, ,
\nonumber \\
\Gamma_{\!\! 22 \, qq \rightarrow tt}^{(1)}&=&C_F\left[2\ln\left(\frac{m_t^2-u}{m_t{\sqrt s}}\right)-1\right]
+\left(C_F-\frac{C_A}{2}\right) \left[2\ln\left(\frac{m_t^2-t}{m_t{\sqrt s}}\right)+L_{\beta}\right] \, ,
\eeqa
where $L_{\beta}$ is given by Eq. (\ref{Lb}).
We, of course, note that this process is similar to $t{\bar t}$ production via the $q{\bar q}$ channel, but a different choice of color basis here leads to a different form for the results.

\subsection{Squark and gluino production}

For squark production via the process $q{\bar q} \rightarrow {\tilde q}{\tilde {\bar q}}$, the soft anomalous dimension is of the same form as for the top-production process $q {\bar q} \rightarrow t{\bar t}$ in Section 8.1.1 (just replace the top-quark mass by the squark mass \cite{NKun04}), with a similar result for the channel $qq \rightarrow {\tilde q}{\tilde q}$ (see also \cite{KM08,BBKKLN}).

For squark production via the process $gg \rightarrow {\tilde q}{\tilde {\bar q}}$, the soft anomalous dimension is of the same form as for the top-production process $gg \rightarrow t{\bar t}$ in Section 8.1.2 (again, just replace the top-quark mass by the squark mass \cite{NKun04}). A modified form of this matrix describes gluino production via the process $q{\bar q} \rightarrow {\tilde g}{\tilde g}$, now using the gluino mass \cite{KM08}. An analogous result describes squark and gluino production via the process $qg \rightarrow {\tilde q} {\tilde g}$ \cite{BBKKLN}.

For gluino production via the process $gg \rightarrow {\tilde g}{\tilde g}$, the color structure is more complicated (the same as for $gg  \rightarrow gg$ in Section 9.5), and the soft anomalous dimension matrix is given in \cite{KM08}.

\mysection{$\Gamma_{\!\! S}$ for jet production and related processes}

In this section we present the soft anomalous dimension matrices for partonic processes involved in jet production \cite{KOS2}; these soft anomalous dimensions are also relevant to related processes such as hadron production.

\subsection{$q{\bar q} \rightarrow q{\bar q}$}

We begin with the quark-antiquark annihilation processes, $q{\bar q} \rightarrow q{\bar q}$.
There are three different types of quark-antiquark processes here, depending on the quark flavors:  $q_j {\bar q}_j \rightarrow q_j {\bar q}_j$, 
$q_j {\bar q}_j \rightarrow q_k {\bar q}_k$, and $q_j {\bar q}_k \rightarrow q_j {\bar q}_k$.

In the $t$-channel singlet-octet color basis
$c_1^{q {\bar q}\rightarrow q {\bar q}}=\delta_{a1}\delta_{b2}$, $c_2^{q {\bar q}\rightarrow q {\bar q}}=T^c_{1a} T^c_{b2}$,
the one-loop soft anomalous dimension matrix is 
\cite{KOS2}
\beq
\Gamma_{\!\! S \, q {\bar q}\rightarrow q {\bar q}}^{(1)}=\left[
                \begin{array}{cc}
                 2C_F \ln(-t/s)  &   -\frac{C_F}{N_c} \ln(-u/s)  \vspace{2mm} \\
                -2\ln(-u/s)    &-\frac{1}{N_c}\ln(-ts/u^2)
                \end{array} \right]\, .
\eeq

At two loops, $\Gamma_{\!\! S \, q {\bar q}\rightarrow q {\bar q}}^{(2)}=K_2 \, \Gamma_{\!\! S \, q {\bar q}\rightarrow q {\bar q}}^{(1)}$.
 
\subsection{$qq \rightarrow qq$ and ${\bar q}{\bar q} \rightarrow {\bar q} {\bar q}$}

Next, we discuss quark-quark scattering processes, 
$qq \rightarrow qq$.
There are two different types of quark-quark processes here, 
depending on the quark flavors: $q_j q_j \rightarrow q_j q_j$ and  $q_j q_k \rightarrow q_j q_k$.

In the $t$-channel octet-singlet color basis
$c_1^{qq \rightarrow qq}=T^c_{1a} T^c_{2b}$, 
$c_2^{qq \rightarrow qq}=\delta_{a1} \delta_{b2}$,
the one-loop soft anomalous dimension matrix is \cite{KOS2}
\beq
\Gamma_{\!\! S \, qq \rightarrow qq}^{(1)}=\left[
                \begin{array}{cc}
                -\frac{1}{N_c}\ln(tu/s^2)+2C_F \ln(-u/s)  &  2 \ln(-u/s) \vspace{2mm} \\
                 \frac{C_F}{N_c} \ln(-u/s)    & 2C_F \ln(-t/s)
                \end{array} \right].
\eeq

At two loops, $\Gamma_{\!\! S \, qq\rightarrow qq}^{(2)}=K_2 \, \Gamma_{\!\! S \, qq \rightarrow qq}^{(1)}$.

The same soft anomalous dimension matrix describes the process with antiquarks, 
${\bar q}{\bar q} \rightarrow {\bar q}{\bar q}$.

\subsection{$q{\bar q} \rightarrow gg$ and $gg \rightarrow q{\bar q}$}

Next, we discuss the processes $q {\bar q} \rightarrow  gg $ and
$gg \rightarrow q {\bar q}$. 

For the process $q{\bar q} \rightarrow gg$,
in the $s$-channel color basis
$c_1^{q {\bar q} \rightarrow gg}=\delta_{ab}\delta_{12}$, $c_2^{q {\bar q} \rightarrow gg}=d^{12c} T^c_{ba}$, 
$c_3^{q {\bar q} \rightarrow gg}=if^{12c} T^c_{ba}$,
the one-loop soft anomalous dimension matrix is \cite{KOS2} 
\beq
\Gamma_{\!\! S \, q {\bar q} \rightarrow gg}^{(1)}=\left[
                \begin{array}{ccc}
                 0  &   0  & \ln(u/t)  \vspace{2mm} \\ 
                 0  &   \frac{C_A}{2}\ln(tu/s^2)    & \frac{C_A}{2}\ln(u/t) \vspace{2mm} \\ 
                 2\ln(u/t)  & \frac{N_c^2-4}{2N_c}\ln(u/t)  & \frac{C_A}{2}\ln(tu/s^2)
                \end{array} \right].
\eeq

At two loops, $\Gamma_{\!\! S \, q {\bar q} \rightarrow gg}^{(2)}=K_2 \, \Gamma_{\!\! S \, q {\bar q} \rightarrow gg}^{(1)}$.

This soft anomalous dimension matrix also describe the time-reversed process 
$gg \rightarrow {\bar q}q$.

\subsection{$q g \rightarrow  q g$ and ${\bar q} g \rightarrow {\bar q} g$}

Here we discuss quark-gluon scattering,
$qg \rightarrow qg$.
In the $t$-channel color basis 
$c_1^{q g \rightarrow  q g}=\delta_{a1}\delta_{b2}$, 
$c_2^{q g \rightarrow  q g}=d^{b2c} T^c_{1a}$, 
$c_3^{q g \rightarrow  q g}=if^{b2c} T^c_{1a}$, 
the one-loop soft anomalous dimension matrix is \cite{KOS2}
\beq
\Gamma_{\!\! S \, q g \rightarrow  q g}^{(1)}=\left[
                \begin{array}{ccc}
                 \left( C_F+C_A \right) \ln(-t/s)  &   0  & \ln(-u/s)  \vspace{2mm} \\ 
                 0  &   C_F \ln(-t/s)+ \frac{C_A}{2} \ln(-u/s)     & \frac{C_A}{2} \ln(-u/s)  
\vspace{2mm} \\
                 2 \ln(-u/s)  & \frac{N_c^2-4}{2N_c} \ln(-u/s)  &  C_F \ln(-t/s)+ \frac{C_A}{2} \ln(-u/s)
                \end{array} \right] \, .
\label{Gammaqgqg}
\eeq

At two loops, $\Gamma_{\!\! S \, q g \rightarrow  q g}^{(2)}=K_2 \, \Gamma_{\!\! S \, q g \rightarrow  q g}^{(1)}$.

This soft anomalous dimension matrix also describes the process 
${\bar q}g \rightarrow {\bar q}g$.

\subsection{$g g \rightarrow  g g $}

Finally, we consider gluon-gluon scattering, $gg \rightarrow gg$.
The color decomposition for this process is by far the most complicated
among $2\rightarrow 2$ processes.  
A complete color basis for the process $gg \rightarrow gg$
is given by the eight color structures~\cite{KOS2}
\beqa 
c_1^{g g \rightarrow  g g}&=&=\frac{i}{4}\left(f^{a b c}
d^{1 2 c} - d^{a b c}f^{1 2 c}\right) \, , \quad 
c_2^{g g \rightarrow  g g}=\frac{i}{4}\left(f^{a b c}
d^{1 2 c} + d^{a b c}f^{1 2 c}\right) \, ,
\nonumber \\ 
c_3^{g g \rightarrow  g g}&=&=\frac{i}{4}\left(f^{a 1 c}
d^{b 2 c}+d^{a 1 c}f^{b 2 c}\right) \, , \quad
c_4^{g g \rightarrow  g g}=P_1(a,b;1,2)=\frac{1}{N_c^2-1}\delta_{a 1} \delta_{b 2} \, ,
\nonumber \\
c_5^{g g \rightarrow  g g}&=&P_{8_S}(a,b;1,2)=\frac{N_c}{N_c^2-4} d^{a1c} d^{b2c} \, , 
\quad c_6^{g g \rightarrow  g g}=P_{8_A}(a,b;1,2)=\frac{1}{N_c} f^{a1c} f^{b2c} \, ,
\nonumber \\
c_7^{g g \rightarrow  g g}&=&P_{10+{\overline{10}}}(a,b;1,2)=    
\frac{1}{2}(\delta_{a b} \delta_{1 2}
-\delta_{a 2} \delta_{b 1})
-\frac{1}{N_c} f^{a1c} f^{b2c} \, ,
\nonumber \\
c_8^{g g \rightarrow  g g}&=&P_{27}(a,b;1,2)=\frac{1}{2}(\delta_{a b} 
\delta_{1 2} +\delta_{a 2} \delta_{b 1})
-\frac{1}{N_c^2-1}\delta_{a 1} \delta_{b 2}
-\frac{N_c}{N_c^2-4} d^{a1c} d^{b2c} \, ,
\label{8x8basis}
\eeqa
where we used the $t$-channel projectors $P$ in the product 
$8 \otimes 8=1+8_S+8_A+10+{\overline{10}}+27$ describing the color content of a set of two gluons. 

The one-loop soft anomalous dimension matrix is~\cite{KOS2}
\beq
\Gamma_{\!\! S \, g g \rightarrow  g g}^{(1)}=\left[\begin{array}{cc}
            \Gamma_{\! \!3 \times 3}^{(1)} & 0_{3 \times 5} \\
              0_{5 \times 3}      & \Gamma_{\!\! 5 \times 5}^{(1)}
\end{array} \right] \, ,
\label{gammagggg}
\eeq
with
\beq
\Gamma_{\!\! 3 \times 3}^{(1)}=\left[
                \begin{array}{ccc}
                  N_c \ln(-t/s)  &   0  & 0  \\
                  0  &  N_c \ln(-u/s) & 0    \\
                  0  &  0  &  N_c\ln(tu/s^2)
                   \end{array} \right]
\eeq
and
\beq
\Gamma_{\!\! 5 \times 5}^{(1)}
=\left[\begin{array}{ccccc}
2N_c \ln(\frac{-t}{s}) & 0 & -2N_c \ln(\frac{-u}{s}) & 0 & 0 \\ 
0  & \frac{N_c}{2}\ln(\frac{-ut^2}{s^3}) & -\frac{N_c}{2}\ln(\frac{-u}{s}) & -N_c\ln(\frac{-u}{s}) & 0 \\ 
-\frac{2N_c}{N_c^2-1}\ln(\frac{-u}{s}) & -\frac{N_c}{2}\ln(\frac{-u}{s}) & \frac{N_c}{2}\ln(\frac{-ut^2}{s^3}) & 0 & -\frac{N_c(N_c+3)}{2(N_c+1)}\ln(\frac{-u}{s}) \\
0 & -\frac{2N_c}{N_c^2-4}\ln(\frac{-u}{s}) & 0 & N_c \ln(\frac{-u}{s}) & -\frac{N_c(N_c+3)}{2(N_c+2)}\ln(\frac{-u}{s})  \\
0 & 0 &-\frac{2}{N_c}\ln(\frac{-u}{s}) &\frac{(N_c+1)(2-N_c)}{N_c}\ln(\frac{-u}{s}) & (N_c+1)\ln(\frac{-u}{s})-2\ln(\frac{-t}{s})
\end{array} \right] 
\nonumber \\ 
\eeq
while at two loops we have $\Gamma_{\!\! S \, gg \rightarrow  gg}^{(2)}=K_2 \, \Gamma_{\!\! S \, gg \rightarrow  gg}^{(1)}$.

\mysection{$\Gamma_{\!\! S}$ for some $2 \rightarrow 3$ processes}

In this section we consider several processes that involve a three-particle final state at leading order.

\subsection{$tqH$, $tqZ$, $tq\gamma$, $tqW$ production}

We begin with processes with three-particle final states involving a top quark produced in association with a Higgs boson or a photon or a $W$ or $Z$ boson \cite{MFNK}.

We begin with the $s$-channel processes
$q(p_a)+{\bar q'}(p_b) \rightarrow t(p_1) +{\bar b}(p_2)+H(p_3)$ 
as well as $q{\bar q'} \rightarrow t{\bar b}Z$, $q{\bar q'} \rightarrow t{\bar b}\gamma$, $q{\bar q} \rightarrow t{\bar b}W^-$, $q{\bar q'} \rightarrow t{\bar q''}W^+$. We define $s=(p_a+p_b)^2$, $t=(p_a-p_1)^2$, and $u=(p_b-p_1)^2$, as before, and further define 
$s'=(p_1+p_2)^2$, $t'=(p_b-p_2)^2$, and $u'=(p_a-p_2)^2$. 
The soft anomalous dimension matrix is identical for all these processes.
We choose $q{\bar q'} \rightarrow t{\bar b}H$ as the specific process, and we use the color basis $c_1^{q{\bar q'}\rightarrow t{\bar b}H}=\delta_{ab} \delta_{12}$ and 
$c_2^{q{\bar q'}\rightarrow t{\bar b}H}=T^c_{ba} T^c_{12}$. Then, the four elements of the soft anomalous dimension matrix, $\Gamma_{\!\! S \, q{\bar q'}\rightarrow t{\bar b}H}$, are given at one loop by \cite{MFNK}
\beqa
\Gamma_{\!\! 11 \, q{\bar q'}\rightarrow t{\bar b}H}^{(1)}&=&C_F \left[\ln\left(\frac{s'-m_t^2}{m_t\sqrt{s}}\right)-\frac{1}{2}\right] \, ,
\nonumber \\
\Gamma_{\!\! 12 \, q{\bar q'}\rightarrow t{\bar b}H }^{(1)}&=&\frac{C_F}{2N_c} \ln\left(\frac{t'(t-m_t^2)}{u'(u-m_t^2)}\right) \, , \quad \quad 
\Gamma_{\!\! 21 \, q{\bar q'}\rightarrow t{\bar b}H }^{(1)}= \ln\left(\frac{t'(t-m_t^2)}{u'(u-m_t^2)}\right) \, ,
\nonumber \\
\Gamma_{\!\! 22 \, q{\bar q'}\rightarrow t{\bar b}H}^{(1)}&=&C_F \left[\ln\left(\frac{s'-m_t^2}{m_t \sqrt{s}}\right)-\frac{1}{2}\right]
-\frac{1}{N_c}\ln\left(\frac{t'(t-m_t^2)}{u'(u-m_t^2)}\right)
+\frac{N_c}{2} \ln\left(\frac{t'(t-m_t^2)}{s(s'-m_t^2)}\right) \, .
\eeqa

We note that this is very similar to $s$-channel single-top production since in both cases we have two colored particles in the final state, the difference being an extra colorless boson in the case here. Thus, the soft anomalous dimension matrices are almost the same, the difference arising from the more complicated kinematics in $tqH$ production; essentially, by replacing $s$ by $s'$, $t$ by $t'$, and $u$ by $u'$ in selected places.   

We continue with the $t$-channel processes
$b(p_a)+q(p_b) \rightarrow t(p_1) +q'(p_2)+H(p_3)$ as well as
$bq \rightarrow tq'Z$, $bq \rightarrow tq'\gamma$, 
$bq \rightarrow tqW^-$, $qq \rightarrow tq'W^+$, 
which have the same soft anomalous dimension matrix.
We define the kinematical variables as above.  
We choose $bq \rightarrow tq'H$ as the specific process, and we use the color basis $c_1^{bq\rightarrow tq'H}=\delta_{a1} \delta_{b2}$ and 
$c_2^{bq\rightarrow tq'H}=T^c_{1a} T^c_{2b}$.
Then, the four elements of the soft anomalous dimension matrix, $\Gamma_{\!\! S \, bq\rightarrow tq'H}$, are given at one loop by \cite{MFNK}
\beqa
{\Gamma}_{\!\! 11 \, bq\rightarrow tq'H}^{(1)}&=&
C_F \left[\ln\left(\frac{t'(t-m_t^2)}{m_t s^{3/2}}\right)-\frac{1}{2}\right] \, ,
\nonumber \\ 
{\Gamma}_{\!\! 12 \, bq\rightarrow tq'H}^{(1)}&=&\frac{C_F}{2N_c} \ln\left(\frac{u'(u-m_t^2)}{s(s'-m_t^2)}\right) \, , \quad \quad 
{\Gamma}_{\!\! 21 \, bq\rightarrow tq'H}^{(1)}= \ln\left(\frac{u'(u-m_t^2)}{s(s'-m_t^2)}\right) \, ,
\nonumber \\
{\Gamma}_{\!\! 22 \, bq\rightarrow tq'H}^{(1)}&=& C_F \left[\ln\left(\frac{t'(t-m_t^2)}{m_t s^{3/2}}\right)-\frac{1}{2}\right]
-\frac{1}{N_c}\ln\left(\frac{u'(u-m_t^2)}{s(s'-m_t^2)}\right) 
+\frac{N_c}{2}\ln\left(\frac{u'(u-m_t^2)}{t'(t-m_t^2)}\right) \, .
\eeqa

We note that this is very similar to $t$-channel single-top production, and the soft anomalous dimension matrices are almost the same, essentially differing by replacing $s$ by $s'$, $t$ by $t'$, and $u$ by $u'$ in selected places. 

At two loops, the soft anomalous dimension matrices for each of these $s$-channel or $t$-channel processes can be written compactly in terms of the corresponding one-loop results \cite{MFNK}, in a way entirely analogous to the $s$-channel and $t$-channel single-top results in Section 7, i.e. as in Eqs. (\ref{Gamma2s}) and (\ref{Gamma2t}).

\subsection{$t{\bar t}H$, $t{\bar t}Z$, $t{\bar t} \gamma$, $t{\bar t}W$ production}

We next consider the processes $q(p_a)+{\bar q}(p_b) \rightarrow t(p_1)+{\bar t}(p_2)+H(p_3)$ as well as $q{\bar q} \rightarrow t{\bar t}Z$, $q{\bar q} \rightarrow t{\bar t} \gamma$, $q{\bar q'} \rightarrow t{\bar t}W^{\pm}$, 
which have the same soft anomalous dimension matrix. We choose $q{\bar q}\rightarrow t{\bar t}H$ as the specific process and use a color tensor basis of $s$-channel singlet and octet exchange,
$c_1^{q{\bar q}\rightarrow t{\bar t}H} = \delta_{ab}\delta_{12}$,  
$c_2^{q{\bar q}\rightarrow t{\bar t}H} =  T^c_{ba} \, T^c_{12}$.
The four matrix elements of $\Gamma_{\!\! S \, q{\bar q}\rightarrow t{\bar t}H}$ are closely related to those for $q{\bar q} \rightarrow t{\bar t}$ \cite{NKGS1,NKGS2,NKnnll} that we presented in Section 8.1.1, and are given at one loop \cite{LLL,KMST} by 
\beqa
\Gamma_{\!\! 11 \, q{\bar q}\rightarrow t{\bar t}H}^{(1)}\!\!\!\!\!&=&\!\!\!\!\!-C_F\left(L_{\beta'}+1\right),
\quad
\Gamma_{\!\! 12 \, q{\bar q}\rightarrow t{\bar t}H}^{(1)}=
\frac{C_F}{2N_c} \Gamma_{\!\! 21 \, q{\bar q}\rightarrow t{\bar t}H}^{(1)} ,
\quad
\Gamma_{\!\! 21 \, q{\bar q}\rightarrow t{\bar t}H}^{(1)}=
\ln\left(\frac{(t-m_t^2)(t'-m_t^2)}{(u-m_t^2)(u'-m_t^2)}\right),
\nonumber \\ 
\Gamma_{\!\! 22 \, q{\bar q}\rightarrow t{\bar t}H}^{(1)}\!\!\!\!\!&=&\!\!\!\!\!\left(C_F-\frac{C_A}{2}\right)
\left[-L_{\beta'}-1+2\ln\left(\frac{(t-m_t^2)(t'-m_t^2)}{(u-m_t^2)(u'-m_t^2)}\right)\right]
+\frac{C_A}{2}\left[\ln\left(\frac{(t-m_t^2)(t'-m_t^2)}{s\, m_t^2}\right)-1\right]
\nonumber \\
\eeqa
where $L_{\beta'}$ is of the form of Eq. (\ref{Lb}) but with $\beta$ replaced by $\beta'=\sqrt{1-4m^2/s'}$.

The processes $gg \rightarrow t{\bar t}H$, $gg \rightarrow t{\bar t}Z$, $gg \rightarrow t{\bar t} \gamma$, have the same  soft anomalous dimension which is a $3\times 3$ matrix of the form of Eq. (\ref{Gsggtt}). The matrix elements are closely related to those for $gg \rightarrow t {\bar t}$ \cite{NKGS2,NKnnll} that we presented in Section 8.1.2. We choose $gg \rightarrow t{\bar t}H$ as the specific process and use the color basis 
$c_1^{gg\rightarrow t{\bar t}H}=\delta^{ab}\,\delta_{12}$, $c_2^{gg\rightarrow t{\bar t}H}=d^{abc}\,T^c_{12}$, $c_3^{gg\rightarrow t{\bar t}H}=i f^{abc}\,T^c_{12}$. 
At one loop we have \cite{LLL,KMST}
\beqa
\Gamma_{\!\! 11 \, gg\rightarrow t{\bar t}H}^{(1)}&=& -C_F\left(L_{\beta'}+1\right)  \, ,
\quad 
\Gamma_{\!\! 13 \, gg\rightarrow t{\bar t}H}^{(1)}= \frac{1}{2}\ln\left(\frac{(t-m_t^2)(t'-m_t^2)}{(u-m_t^2)(u'-m_t^2)}\right) \, , 
\nonumber \\
\Gamma_{\!\! 22 \, gg\rightarrow t{\bar t}H}^{(1)}&=& \left(C_F-\frac{C_A}{2}\right) \left(-L_{\beta'}-1\right)
+\frac{C_A}{2}\left[\frac{1}{2}\ln\left(\frac{(t-m_t^2)(t'-m_t^2)(u-m_t^2)(u'-m_t^2)}{s^2\, m_t^4}\right)-1\right] \, ,
\nonumber \\
\Gamma_{\!\! 31 \, gg\rightarrow t{\bar t}H}^{(1)}&=&2 \, \Gamma_{\!\! 13 \, gg\rightarrow t{\bar t}H}^{(1)} \, , \quad  \Gamma_{\!\! 23 \, gg\rightarrow t{\bar t}H}^{(1)}=\frac{C_A}{2} \, \Gamma_{\!\! 13 \, gg\rightarrow t{\bar t}H}^{(1)} \, , \quad 
 \Gamma_{\!\! 32 \, gg\rightarrow t{\bar t}H}^{(1)}=\frac{(N_c^2-4)}{2N_c} \, \Gamma_{\!\! 13 \, gg\rightarrow t{\bar t}H}^{(1)} \, .
\eeqa

\subsection{$q{\bar q}g$, $Q{\bar Q}g$, and $ggg$ final states}

Soft anomalous dimension matrices at one loop for processes with three colored particles in the final state have appeared in Refs. \cite{MS,ES}. 

The soft anomalous dimension for the process $q{\bar q} \rightarrow q{\bar q}g$ is a $4\times 4$ matrix, for the process $gg \rightarrow q{\bar q}g$ it is an $11\times 11$ matrix, and for the process $gg \rightarrow ggg$ it is a $22\times 22$ matrix, with details given in Ref. \cite{MS}.

Results for related processes involving heavy quarks were given in Ref. \cite{ES}. The soft anomalous dimension for the process 
$q{\bar q} \rightarrow Q{\bar Q}g$ is again a $4\times 4$ matrix, and for the process $gg \rightarrow Q{\bar Q}g$ it is again an $11\times 11$ matrix, with details given in Ref. \cite{ES}.

\mysection{Summary and Conclusions}

Soft-gluon resummation provides a powerful method to calculate large, and often dominant, higher-order corrections in perturbative cross sections (see Ref. \cite{NKtoprev} for numerical results for many processes). Soft anomalous dimensions are essential in performing resummation beyond leading-logarithm accuracy and, in general, they are matrices in the space of color exchanges. 

One-loop results for soft anomalous dimensions are available for virtually all $2 \to 2$ processes as well as many $2 \to 3$ processes. Two-loop results and even three-loop results are also known for many $2 \to 2$ processes and some $2 \to 3$ ones. We have reviewed these results using a consistent approach and terminology for all of them. We have provided comprehensive and detailed expressions for a large number of $2 \to 2$ processes involving single-top and top-pair production, electroweak-boson and Higgs production, jet production, and other SM and BSM processes. 

We have also provided results for soft anomalous dimensions for a number of $2 \to 3$ processes involving the production of single top quarks or top-antitop pairs in association with electroweak or Higgs bosons, and discussed processes with three final-state colored particles.

These results can be used, and have been used, for performing resummation and for calculating soft-gluon corrections at higher orders for a very large number of processes.

\mysection*{Acknowledgements}
This material is based upon work supported by the National Science Foundation under Grant No. PHY 1820795.


\begin{thebibliography}{99}

\bibitem{GS87}
G. Sterman, Nucl. Phys. B {\bf 281}, 310 (1987). 

\bibitem{CT89}
S. Catani and L. Trentadue, Nucl. Phys. B {\bf 327}, 323 (1989).

\bibitem{NKGS1}
N. Kidonakis and G. Sterman,  Phys. Lett. B {\bf 387}, 867 (1996). 

\bibitem{CLS97}
H. Contopanagos, E. Laenen, and G. Sterman, Nucl. Phys. B {\bf 484}, 303 (1997) [hep-ph/9604313]. 

\bibitem{NKGS2}
N. Kidonakis and G. Sterman, Nucl. Phys. B {\bf 505}, 321 (1997) [hep-ph/9705234].

\bibitem{KOS1}
N. Kidonakis, G. Oderda, and G. Sterman, Nucl. Phys. B {\bf 525}, 299 (1998) [hep-ph/9801268]. 

\bibitem{KOS2}
N. Kidonakis, G. Oderda, and G. Sterman, Nucl. Phys. B {\bf 531}, 365 (1998) [hep-ph/9803241].

\bibitem{LOS}
E. Laenen, G. Oderda, and G. Sterman, Phys. Lett. B {\bf 438}, 173 (1998) [hep-ph/9806467].

\bibitem{ADS} 
S.M. Aybat, L.J. Dixon, and G. Sterman, Phys. Rev. Lett. {\bf 97}, 072001 (2006) [hep-ph/0606254]. 

\bibitem{HRSV}
P. Hinderer, F. Ringer, G. Sterman, and W. Vogelsang, Phys. Rev. D {\bf 99}, 054019 (2019) [arXiv:1812.00915].

\bibitem{MFNK}
M. Forslund and N. Kidonakis, Phys. Rev. D {\bf 102}, 034006 (2020) [arXiv:2003.09021].

\bibitem{NKtoprev}
N. Kidonakis, Int. J. Mod. Phys. A {\bf 33}, 1830021 (2018) [arXiv:1806.03336].

\bibitem{NKBP}
N. Kidonakis and B.D. Pecjak, Eur. Phys. J. C {\bf 72}, 2084 (2012) [arXiv:1108.6063].

\bibitem{GW}
D.J. Gross and F. Wilczek, Phys. Rev. Lett. {\bf 30}, 1343 (1973).

\bibitem{HDP}
H.D. Politzer, Phys. Rev. Lett. {\bf 30}, 1346 (1973).

\bibitem{beta1a}
W.E. Caswell, Phys. Rev. Lett. {\bf 33}, 244 (1974). 

\bibitem{beta1b}
D.R.T. Jones, Nucl. Phys. B {\bf 75}, 531 (1974).

\bibitem{beta1c}
E. Egorian and O.V. Tarasov, Teor. Mat. Fiz. {\bf 41}, 26 (1979), 
Theor. Math. Phys. {\bf 41}, 863 (1979).

\bibitem{beta2a}
O.V. Tarasov, A.A. Vladimirov, and A.Yu. Zharkov, Phys. Lett. B {\bf 93}, 429 (1980). 

\bibitem{beta2b}
S.A. Larin and J.A.M. Vermaseren, Phys. Lett. B {\bf 303}, 334 (1993) [hep-ph/9302208].

\bibitem{beta3}
T. van Ritbergen, J.A.M. Vermaseren, and S.A. Larin, Phys. Lett. B {\bf 400}, 379 (1997) [hep-ph/9701390].

\bibitem{beta4}
F. Herzog, B. Ruijl, T. Ueda, J.A.M. Vermaseren, and A. Vogt, JHEP {\bf 1702}, 090 (2017) [arXiv:1701.01404].

\bibitem{NK2loop}
N. Kidonakis, Phys. Rev. Lett. {\bf 102}, 232003 (2009) [arXiv:0903.2561].

\bibitem{KT82}
J. Kodaira and L. Trentadue, Phys. Lett. {\bf 112B}, 66 (1982).

\bibitem{MVV04}
S. Moch, J.A.M. Vermaseren, and A. Vogt, Nucl. Phys. B {\bf 688}, 101 (2004) [hep-ph/0403192].

\bibitem{HKM}
J.M. Henn, G.P. Korchemsky, and B. Mistlberger, JHEP {\bf 2004}, 018 (2020) [arXiv:1911.10174].

\bibitem{MPS}
A. von Manteuffel, E. Panzer, and R.M. Schabinger, Phys. Rev. Lett. {\bf 124}, 162001 (2020) [arXiv:2002.04617].

\bibitem{MA05}
S. Moch and A. Vogt, Phys. Lett. B {\bf 631}, 48 (2005) [hep-ph/0508265].

\bibitem{FRS1}
E.G. Floratos, D.A. Ross, and C.T. Sachrajda, Nucl. Phys. B {\bf 129}, 66 (1977) [(E) B {\bf 139}, 545 (1978)].

\bibitem{FRS2}
E.G. Floratos, D.A. Ross, and C.T. Sachrajda, Nucl. Phys. B {\bf 152}, 493 (1979).

\bibitem{GALY79}
A. Gonzalez-Arroyo, C. Lopez, and F.J. Yndurain, Nucl. Phys. B {\bf 153}, 161 (1979).
 
\bibitem{GFP80}
G. Curci, W. Furmanski, and R. Petronzio, Nucl. Phys. B {\bf 175}, 27 (1980).

\bibitem{FP80}
W. Furmanski and R. Petronzio, Phys. Lett. {\bf 97B}, 437 (1980).

\bibitem{MVV05}
S. Moch, J.A.M. Vermaseren, and A. Vogt, Nucl. Phys. B {\bf 726}, 317 (2005) [hep-ph/0506288].

\bibitem{NKun04}
N. Kidonakis, Int. J. Mod. Phys. A {\bf 19}, 1793 (2004) [hep-ph/0303186].

\bibitem{BN}
T. Becher and M. Neubert, Phys. Rev. Lett. {\bf 102}, 162001 (2009) [(E) {\bf 111}, 199905 (2013)] [arXiv:0901.0722]. 

\bibitem{GM}
E. Gardi and L. Magnea, JHEP {\bf 0903}, 079 (2009) [arXiv:0901.1091]. 

\bibitem{ADG} 	
O. Almelid, C. Duhr, and E. Gardi, Phys. Rev. Lett. {\bf 117}, 172002 (2016) [arXiv:1507.00047].

\bibitem{NKNNNLO}
N. Kidonakis, Phys. Rev. D {\bf 73}, 034001 (2006) [hep-ph/0509079].

\bibitem{AMP}
A.M. Polyakov, Nucl. Phys. B {\bf 164}, 171 (1980).

\bibitem{BNS}
R.A. Brandt, F. Neri, and M. Sato, Phys. Rev. D {\bf 24}, 879 (1981).

\bibitem{IKR}
S.V. Ivanov, G.P. Korchemsky, and A.V. Radyushkin, 
Yad. Fiz. {\bf 44}, 230 (1986) [Sov. J. Nucl. Phys. {\bf 44}, 145 (1986)].  

\bibitem{KR}
G.P. Korchemsky and A.V. Radyushkin, Phys. Lett. B {\bf 171}, 459 (1986). 

\bibitem{GHKM}
A. Grozin, J.M. Henn, G.P. Korchemsky, and P. Marquard, 
Phys. Rev. Lett. {\bf 114}, 062006 (2015) [arXiv:1409.0023]. 

\bibitem{NK3loopcusp}
N. Kidonakis, Int. J. Mod. Phys. A {\bf 31}, 1650076 (2016) [arXiv:1601.01666].

\bibitem{ERJV}
E. Remiddi and J.A.M. Vermaseren, Int. J. Mod. Phys. A {\bf 15}, 725 (2000) [hep-ph/9905237].

\bibitem{NKRG}
N. Kidonakis and R.J. Gonsalves, Phys. Rev. D {\bf 87}, 014001 (2013) [arXiv:1201.5265].

\bibitem{NKst}
N. Kidonakis, Phys. Rev. D {\bf 74}, 114012 (2006) [hep-ph/0609287].

\bibitem{NKsch}
N. Kidonakis, Phys. Rev. D {\bf 81}, 054028 (2010) [arXiv:1001.5034].

\bibitem{NKtop3loop}
N. Kidonakis, Phys. Rev. D {\bf 99}, 074024 (2019) [arXiv:1901.09928].

\bibitem{NKtch}
N. Kidonakis,  Phys. Rev. D {\bf 83}, 091503 (2011) [arXiv:1103.2792].

\bibitem{NKtWH}
N. Kidonakis, Phys. Rev. D {\bf 82}, 054018 (2010) [arXiv:1005.4451].

\bibitem{ABNK}
A. Belyaev and N. Kidonakis, Phys. Rev. D {\bf 65}, 037501 (2002) [hep-ph/0102072].

\bibitem{NKAB}
N. Kidonakis and A. Belyaev, JHEP {\bf 0312}, 004 (2003) [hep-ph/0310299].

\bibitem{NKEM}
N. Kidonakis and E. Martin, Phys. Rev. D {\bf 90}, 054021 (2014) [arXiv:1404.7488].

\bibitem{FNPY1}
A. Ferroglia, M. Neubert, B.D. Pecjak, and L.L. Yang, Phys. Rev. Lett. {\bf 103}, 201601 (2009) [arXiv:0907.4791].

\bibitem{FNPY2}
A. Ferroglia, M. Neubert, B.D. Pecjak, and L.L. Yang, JHEP {\bf 0911}, 062 (2009) [arXiv:0908.3676].

\bibitem{NKnnll}
N. Kidonakis, Phys. Rev. D {\bf 82}, 114030 (2010) [arXiv:1009.4935].

\bibitem{ELSM}
E. Laenen and S. Moch, Phys. Rev. D {\bf 59}, 034027 (1999) [hep-ph/9809550].

\bibitem{KM08}
A. Kulesza and L. Motyka, Phys. Rev. Lett. {\bf 102}, 111802 (2009) [arXiv:0807.2405].

\bibitem{BBKKLN}
W. Beenakker, S. Brensing, M. Kramer, A. Kulesza, E. Laenen, and I. Niessen, 
JHEP {\bf 0912}, 041 (2009) [arXiv:0909.4418].

\bibitem{LLL}
H.T. Li, C.S. Li, and S.A. Li, Phys. Rev. D {\bf 90}, 094009 (2014) [arXiv:1409.1460].

\bibitem{KMST}
A. Kulesza, L. Motyka, T. Stebel, and V. Theeuwes, 
JHEP {\bf 1603}, 065 (2016) [arXiv:1509.02780]. 

\bibitem{MS}
M. Sjodahl, JHEP {\bf 0812}, 083 (2008) [arXiv:0807.0555].

\bibitem{ES}
E. Szarek, Acta Phys. Polon. B {\bf 49}, 1839 (2018) [arXiv:1809.00384].

\end{thebibliography}
\end{document}